\documentclass[aps,prl,twocolumn,superscriptaddress,showpacs,longbibliography]{revtex4-1}

\usepackage{amsmath}
\usepackage{graphicx}
\usepackage{amsfonts}
\usepackage{verbatim}
\usepackage{amssymb}
\usepackage{color}
\usepackage{dsfont}
\usepackage{amsmath} 
\usepackage{hyperref}
\hypersetup{colorlinks = true, urlcolor = blue, linkcolor = blue, citecolor = blue}

\newcommand{\bmat}{\left(\begin{matrix}}
\newcommand{\emat}{\end{matrix}\right)}
\newcommand{\kfl}{k_{F_{n,y}}L_y}
\newcommand{\kfx}{k_{F_{n,x}}}
\newcommand{\kfy}{k_{F_{n,y}}}
\newcommand{\dkfla}{\delta k_{F_{n,y}}L_y}
\newcommand{\dkfl}{\delta k_{F_{n,y}}L_y}

\newcommand{\sinkm}{\sin\left(\frac{k_{F_{n,y}}L_y -\theta_n}{2}\right)}
\newcommand{\sinkp}{\sin\left(\frac{k_{F_{n,y}}L_y +\theta_n}{2}\right)}
\newcommand{\coskm}{\cos\left(\frac{k_{F_{n,y}}L_y -\theta_n}{2}\right)}
\newcommand{\coskp}{\cos\left(\frac{k_{F_{n,y}}L_y +\theta_n}{2}\right)}
\newcommand{\wsc}{W_{\mathrm{SC}}}
\newcommand{\sgn}{\operatorname{sgn}}
\newcommand{\sinc}{\operatorname{sinc}}

\begin{document}

\title{Topological superconductivity in planar Josephson junctions: Narrowing down to the nanowire limit}

\author{F. Setiawan}\email{setiawan@uchicago.edu}
\affiliation{The James Franck Institute and Department of Physics, University of Chicago, Chicago, Illinois 60637, USA}
\author{Ady Stern}
\affiliation{Department of Condensed Matter Physics, Weizmann Institute of Science, Rehovot 76100, Israel}
\author{Erez Berg}
\affiliation{The James Franck Institute and Department of Physics, University of Chicago, Chicago, Illinois 60637, USA}
\affiliation{Department of Condensed Matter Physics, Weizmann Institute of Science, Rehovot 76100, Israel}
\date{\today}

\begin{abstract}
We theoretically study topological planar Josephson junctions (JJs) formed from spin-orbit-coupled two-dimensional electron gases (2DEGs) proximitized by two superconductors and subjected to an in-plane magnetic field $B_\parallel$. Compared to previous studies of topological superconductivity in these junctions, here we consider the case where the superconducting leads are narrower than the superconducting coherence length. In this limit the system may be viewed as a proximitized multiband wire, with an additional knob being the phase difference $\phi$ between the superconducting leads.
A combination of mirror and time-reversal symmetry may put the system  into the class BDI. Breaking this symmetry changes the symmetry class to class D. The class D phase diagram depends strongly on $B_{\parallel}$ and chemical potential, with a weaker dependence on $\phi$ for JJs with narrower superconducting leads. In contrast, the BDI phase diagram depends strongly on both $B_\parallel$ and $\phi$.  Interestingly, the BDI phase diagram has a ``fan"-shaped region with phase boundaries which move away from $\phi = \pi$ linearly with $B_\parallel$. The number of distinct phases in the fan increases with increasing chemical potential. We study the dependence of the JJ's critical current on $B_\parallel$, and find that minima in the critical current indicate first-order phase transitions in the junction only when the spin-orbit coupling strength is small. In contrast to the case of a JJ with wide leads, in the narrow case these transitions are not accompanied by a change in the JJ's topological index. Our results, calculated using realistic experimental parameters, provide guidelines for present and future searches for topological superconductivity in JJs with narrow leads, and are particularly relevant to recent experiments on InAs 2DEGs proximitized by narrow Al superconducting leads [A. Fornieri \textit{et al.},  \href{https://www.nature.com/articles/s41586-019-1068-8}{Nature (London) \textbf{569}, 89 (2019)}].
\end{abstract}

\maketitle
Majorana zero modes (MZMs)~\cite{alicea2012new,Elliott2015Colloquium,leijnse2012_introduction,Beenakker2013Search,leijnse2012_introduction,stanescu2013majorana,sarma2015majorana,beenakker2016road,aguado2017majorana,lutchyn2018majorana} are not only of fundamental interest but can also be used as the building blocks for a fault tolerant quantum computation~\cite{kitaev2003fault,Nayak2008nonabelian}. These MZMs exist in the vortex core of two-dimensional (2D) topological superconductors (TSCs)~\cite{Ivanov2001Non,Read2000Paired} or at the edge of 1D TSCs~\cite{Kitaev2001Unpaired}. The theoretical proposals on TSCs~\cite{Kitaev2001Unpaired,Read2000Paired,Lutchyn2010Majorana,Oreg2010Helical,Fu2008Superconducting,Choy2011Majorana,Nadj2013Proposal,Brydon2015Topological} have triggered a tremendous amount of experimental effort to realize TSCs in different platforms ranging from 1D nanowires~\cite{Mourik2012Signatures,Rokhinson2012fractional,Deng2012Anomalous,Das2012Zero,Churchill2013Superconductor,Finck2013Anomalous,albrecht2016exponential,gul2018ballistic,chen2017experimental,deng2016majorana,Suominen2017Zero,Nichele2017Scaling,zhang2018quantized,zhang2017ballistic,Sestoft2018Engineering,Deng2018Nonlocality,laroche2017observation,van2018observation,de2018electric,grivnin2018concomitant,vaitiekenas2018flux}, topological insulators~\cite{Xu2015Experimental,HaoHua2016Majorana}, and ferromagnetic atomic chains~\cite{nadj2014observation,feldman2017high,jeon2017distinguishing,pawlak2016probing}. Recently, a two-dimensional electron gas (2DEG) with strong spin-orbit coupling (SOC) and proximitized by two spatially separated superconductors (SCs), thus forming a Josephson junction (JJ), was  proposed as a new platform to engineer TSCs~\cite{Pientka2017Topological,Hell2017Two}. Compared to the other setups, this system has the advantage of being able to be tuned into a TSC by changing not only the strength of the applied magnetic field but also the superconducting phase difference $\phi$ across the JJ~\cite{Pientka2017Topological,Hell2017Two}. Recent experiments~\cite{Antonio2018Evidence,Ren2018Topological} using this setup have observed some evidence of Zeeman- and phase-tunable topological superconductivity in form of zero-bias conductance peaks.

In the presence of a symmetry which is a product of the mirror and time-reversal symmetries~\cite{Pientka2017Topological,Hell2017Two}, the topological planar JJ belongs to the symmetry class BDI in the tenfold classification~\cite{ryu2010topological,kitaev2009periodic}, characterized by a $\mathbb{Z}$ topological invariant $Q_{\mathbb{Z}}$. This invariant corresponds to the number of MZMs at the junction's end. Breaking this symmetry changes the symmetry class to D with a $\mathbb{Z}_2$ index. For JJs with SCs whose width $W_{\mathrm{SC}}$ is much larger than the coherence length $\xi$ (as studied in Refs.~\cite{Pientka2017Topological,Hell2017Two,Liu2018Long}), the class BDI and D phase diagrams have a weak dependence on the chemical potential but depend strongly on both the Zeeman field and $\phi$~\cite{Pientka2017Topological}. Moreover, if $\phi$ is not externally controlled, then as the Zeeman field is varied the system undergoes a first-order topological phase transition (TPT) where the phase of the ground state jumps from $\phi \approx 0$ (trivial) to $\phi \approx \pi$ (topological) or vice versa. This phase jump~\cite{Pientka2017Topological,Zyuzin2016Josephson,Alidoust2017Spontaneous,Alidoust2018Strain} is accompanied by a minimum in the critical current which can be used as an experimental probe for the TPT. 

\begin{figure}
\centering
{\includegraphics[width = \linewidth]{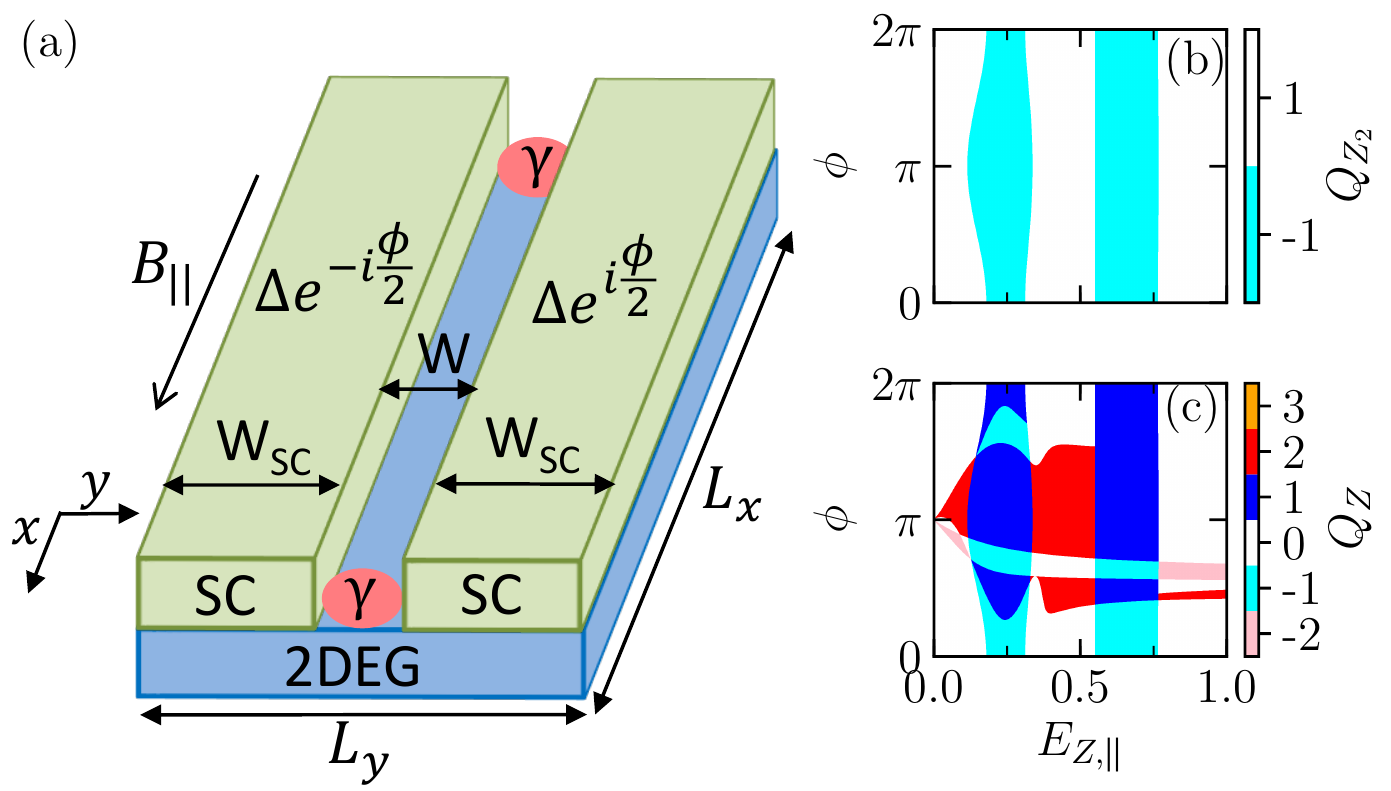}}
\caption{(a) A JJ made of two narrow SC leads in contact with a 2DEG. By applying an in-plane Zeeman field $E_{Z,||} = g\mu_B B_{||}/2$ parallel to the JJ and a superconducting phase difference $\phi$, the system can be tuned into a TSC supporting MZMs $\gamma$. (b) Class D and (c) class BDI phase diagrams as functions of $E_{Z,||}$ and $\phi$. Regions with odd and even $Q_\mathbb{Z}$ topological index in the class BDI [(c)] correspond respectively to the topological ($Q_{\mathbb{Z}_2} = -1$) and trivial ($Q_{\mathbb{Z}_2} = 1$) regions of the class D [(b)]. The phase diagrams are obtained from numerical simulations performed using the Kwant package~\cite{groth2014kwant} of a tight-binding version of Eq.~\eqref{eq:HSM}~(see Sec.~\ref{sec:tbh} of Ref.~\cite{suppl}). 
The parameters used correspond to the experimental parameters of recent experiments on InAs 2DEGs~\cite{Antonio2018Evidence}, i.e., $m^* = 0.026m_e$, $\alpha = 0.1$ eV{\AA}, $\mu = 0.6$ meV, $\Delta = 0.15$ meV [$\xi = \hbar v_F/(\pi\Delta)$ = 126 nm], $W = 80$ nm, and $W_{\mathrm{SC}} = 160$ nm.}
\label{fig:SNS_schematic}
\end{figure}

Motivated by recent experiments on InAs 2DEGs proximitized by narrow Al SCs~\cite{Antonio2018Evidence}, in this Rapid Communication we study the topological superconductivity in planar JJs with narrow SCs ($W_{\mathrm{SC}} < \xi$), (see Fig. \ref{fig:SNS_schematic}). We further examine the relation between this system and a 1D multiband nanowire TSC~\cite{Lutchyn2011Search,Stanescu2011Majorana}. We establish numerically and analytically that the class D phase diagram depends strongly on the in-plane magnetic field $B_{\parallel}$ applied along the junction, but only weakly on the superconducting phase difference $\phi$. This is due to the presence of multiple normal reflections that originate from the interfaces of the SC leads with the vacuum.
Furthermore, the normal reflections make the phase diagram more sensitive to the 2DEG chemical potential. In contrast, the BDI phase diagram is strongly dependent on both $B_\parallel$ and $\phi$. Crucially, it exhibits a ``fan"-shaped region emerging from $\phi = \pi$ at $B_\parallel = 0$ where the BDI phase boundary lines diverge away from $\phi = \pi$ linearly with $B_\parallel$. The number of distinct BDI phases in the fan increases with increasing chemical potential $\mu$ as there are more occupied subbands for a larger $\mu$.
In addition, the critical current through the junction has minima as a function of $B_\parallel$. These minima correspond to discontinuous transitions of the value of $\phi$ that minimizes the free energy. However, unlike the case of wide SC leads, here these transitions are not necessarily accompanied by a change in the topological index.

The Hamiltonian for the planar JJs [Fig.~\ref{fig:SNS_schematic}(a)] in the Nambu basis $\Psi_{k_x} = (\psi_{k_x,\uparrow},\psi_{k_x,\downarrow},\psi_{-k_x,\downarrow}^\dagger,-\psi_{-k_x,\uparrow}^\dagger)^T$ is $H = \frac{1}{2}\int dk_x \int dy \Psi_{k_x}^\dagger(y) \mathcal{H}_{k_x}(y) \Psi_{k_x}(y)$ where
\begin{align}\label{eq:HSM}
\mathcal{H}_{k_x}(y) &=  \left(\frac{\hbar^2(k_x^2 - \partial_y^2)}{2m^*} - \mu \right)\tau_z + \alpha (k_x \sigma_y + i\partial_y\sigma_x)\tau_z\nonumber\\
&\hspace{2 cm} + E_{Z,||} \sigma_x + \Delta(y)\tau_+ + \Delta^*(y)\tau_-,
 \end{align}
with $\psi_{k_x,\uparrow/\downarrow}(y)$ being the annihilation operator of an electron with spin $\uparrow/\downarrow$ and momentum $k_x$. Throughout most of this Rapid Communication, we assume the JJ to be infinitely long. The Pauli matrices $\boldsymbol{\tau}$ and $\boldsymbol{\sigma}$ act in particle-hole and spin spaces, respectively, and $\tau_{\pm} = (\tau_x \pm i \tau_y)/2$.  Here, $m^*$ is the effective electron mass in the 2DEG, $\mu$ is the chemical potential, $\alpha$ is the Rashba SOC strength, and $E_{Z,||} = g \mu_B B_\parallel/2$ is the Zeeman energy due to the applied in-plane magnetic field $B_\parallel$. The proximity-induced pairing potential in the 2DEG is [see Fig.~\ref{fig:SNS_schematic}(a)]
\begin{equation}\label{eq:deltay}
\Delta(y) = \begin{cases}
\Delta e^{-i\phi/2} & \text{for $-(W_{\mathrm{SC}}+ W/2)< y < -W/2$},\\
0 & \text{for $-W/2 < y < W/2$},\\
\Delta e^{i\phi/2} & \text{for $W/2 < y < W_{\mathrm{SC}}+W/2$}.
\end{cases}
\end{equation}

The Hamiltonian in Eq.~\eqref{eq:HSM} anticommutes with the particle-hole operator $P = \sigma_y\tau_yK$, where $K$ denotes complex conjugation. When $E_{Z,||}=0$ and  $\phi = 0$ or $\pi$, the Hamiltonian commutes the standard time-reversal operator $T = -i\sigma_y K$ (where $T^2 = -1$) and thus it belongs to the symmetry class DIII~\cite{ryu2010topological,kitaev2009periodic}. It also commutes with the mirror operator along the $x$-$z$ plane, i.e., $M_y = -\sigma_y \times (y \rightarrow -y)$.
While the $T$ and $M_y$ symmetries are broken when $E_{Z,||}\ne 0$ and/or $\phi \ne 0, \pi$, the Hamiltonian remains invariant under the product $\widetilde{T} = M_yT = i K \times (y \rightarrow -y)$~\cite{Pientka2017Topological}. Since $\widetilde{T}^2 = 1$, the system belongs to the class BDI.  The presence of $\widetilde{T}$ and $P$ symmetries implies that the Hamiltonian anticommutes with the chirality operator $\widetilde{C} = -iP\widetilde{T} = M_y\tau_y$. Breaking the $\widetilde{T}$ symmetry reduces the symmetry class from BDI to D.

To obtain the phase diagrams, we calculate the topological invariant following Ref.~\cite{Tewari2012Topological}. Since the chirality operator obeys $\widetilde{C}^2 =1$, it has eigenvalues $\pm 1$. In the basis where $\widetilde{C}$ is diagonal, the Hamiltonian is block off-diagonal (since $\{\widetilde{C},H\} = 0$). The $\mathbb{Z}$ topological invariant ($Q_\mathbb{Z}$) of the class BDI is calculated from the phase of the determinant of the off-diagonal part. The winding of this phase from $k_x = 0$ to $k_x = 2\pi$ gives $Q_\mathbb{Z}$. The $\mathbb{Z}_2$ index of class D is simply the parity of $Q_\mathbb{Z}$, i.e., $Q_{\mathbb{Z}_2} = (-1)^{Q_{\mathbb{Z}}}$~\cite{Tewari2012Topological,Kitaev2001Unpaired}.

\begin{figure*}[t!]
\centering
{\includegraphics[width = 0.7\linewidth]{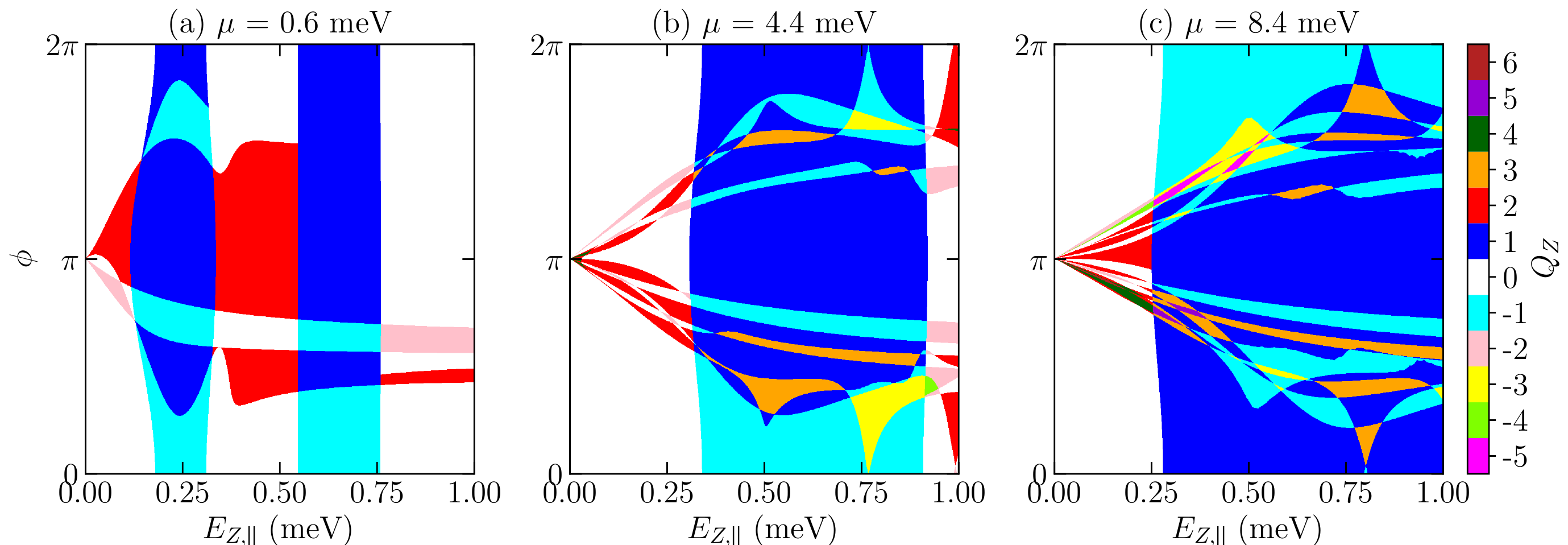}}
\caption{BDI phase diagrams as functions of Zeeman field $E_{Z,||}$ and superconducting phase difference $\phi$ for different chemical potentials: (a) $\mu = 0.6$ meV, (b) $\mu = 4.4$ meV, and (c) $\mu = 8.4$ meV. The phase boundary lines inside the fan-shaped region emanate from $\phi = \pi$ and $E_{Z,||} = 0$ with slopes which are linearly proportional to $E_{Z,||}$ and decrease with increasing $\mu$. The parameters used are the same as in Fig.~\ref{fig:SNS_schematic}.
}\label{fig:phasediagram}
\end{figure*}

Figure~\ref{fig:SNS_schematic}(b) shows the class D phase diagram of a JJ with narrow leads ($W_{\mathrm{SC}} \lesssim \xi$), calculated numerically. The phase diagram shows a sequence of TPTs from the trivial ($Q_{\mathbb{Z}_2} = 1$) to topological ($Q_{\mathbb{Z}_2} = -1$) phases. Contrary to the case of wide SC leads~\cite{Pientka2017Topological}, the phase  boundaries depend only moderately on $\phi$. As $W_{\mathrm{SC}}$ becomes smaller, the strength of normal reflections from the SC edges increases resulting in a weaker dependence of the class D phase boundaries on $\phi$~\footnote{It was shown recently in Ref.~\cite{setiawan2019full} that the chemical potential mismatch between the parent superconductors and 2DEG can also increase the strength of normal reflections which in turn weakens the dependence of the class D phase boundaries on $\phi$.} and the physics crosses over to that of the 1D multiband nanowire TSC~\cite{Lutchyn2011Search,Stanescu2011Majorana}. 

The BDI phase diagram [Fig.~\ref{fig:SNS_schematic}(c)], on the other hand, depends strongly on both $E_{Z,||}$ and $\phi$. For $E_{Z,||}=0$, the BDI topological invariant is $Q_\mathbb{Z}=0$, except at $\phi=\pi$ where the gap closes. As $E_{Z,||}$ increases, the gap closing point expands into a fan-shaped region containing phases with different values of $Q_\mathbb{Z}$.

These features of the phase diagram can be understood qualitatively as follows. Phase transitions where $Q_{\mathbb{Z}_2}$ changes require gap closings at $k_x=0$, while transitions with an even change in $Q_{\mathbb{Z}}$ occur as a consequence of gap closings at the Fermi wave vector, $k_x = \pm k_F$. In the limit where $\xi \ll W_\text{SC}$, the system can be treated as a multiband nanowire~\cite{Lutchyn2011Search,Stanescu2011Majorana}, with an induced gap that is smaller than the energy spacing between subbands. For generic values of $\mu$, the spectrum at $k_x=0$ is gapped for all $\phi$, and therefore the phase diagram depends only weakly on $\phi$. This situation changes at special values of $\mu$ and $E_{Z,||}$, where the chemical potential enters a new subband~(see Sec.~\ref{sec:classDphasediagram} of Ref.~\cite{suppl} for details). Independently of $\mu$, a gap closing occurs at $k_x = \pm k_F$ for $\phi=\pi$ and $E_{Z,||}=0$. This gap closing occurs as a consequence of the mirror symmetry, where the effective induced gap, which is a spatial average of the gap of two symmetric SC leads, vanishes for $\phi=\pi$ and $E_{Z,||}=0$.

\begin{figure}
\centering
{\includegraphics[width = \linewidth]{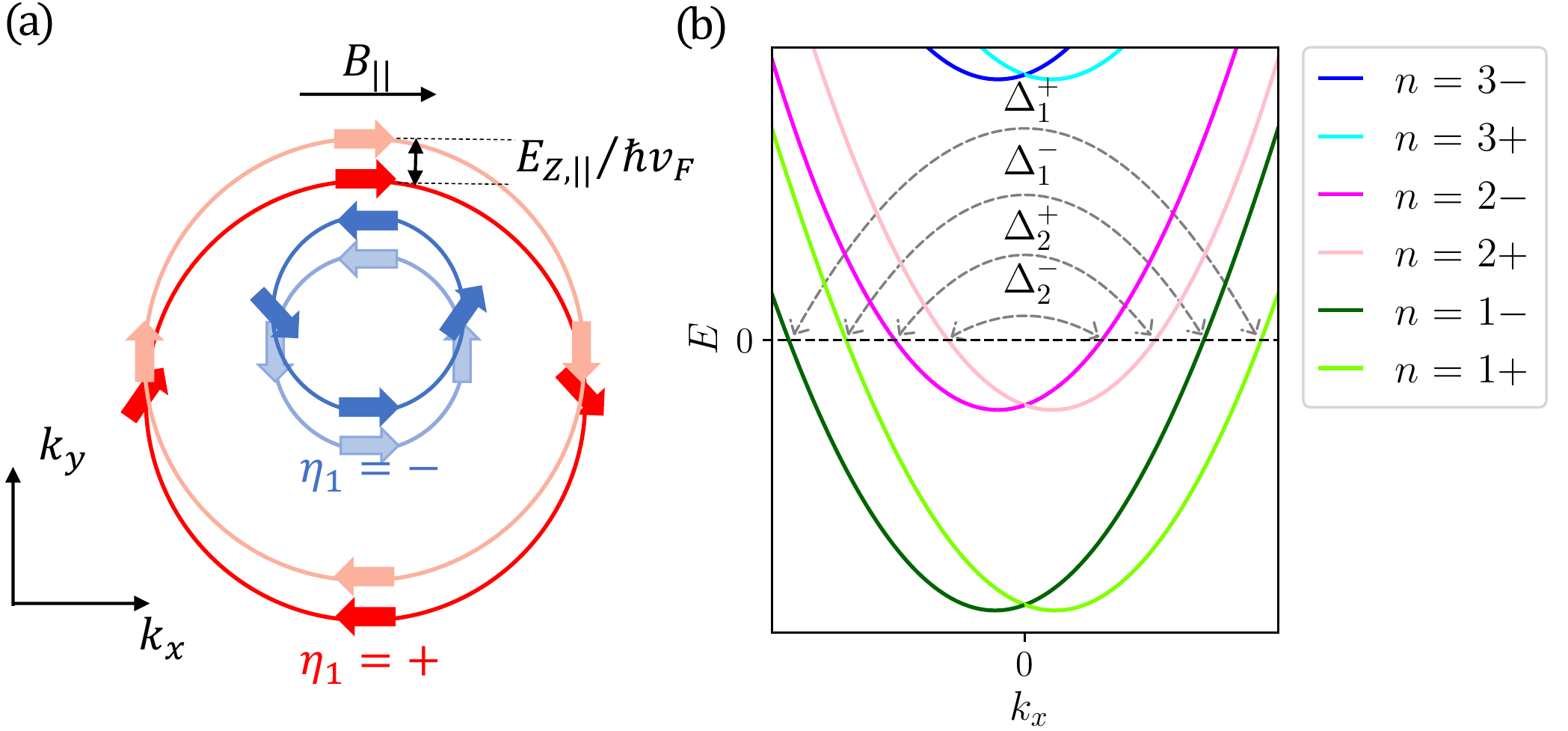}}
\caption{(a) Fermi surfaces of the 2DEG. An applied magnetic field along $x$ shifts the two spin-orbit split Fermi surfaces (labeled by $\eta_1 = \pm$ for the outer and inner Fermi surface) oppositely along $k_y$.  The arrows show the spin orientation on the Fermi surfaces. The Zeeman field tilts the spin-orientation angle towards its direction. (b) Energy spectrum of an infinitely long 2DEG with a finite width. Each $n$th band consists of two subbands labeled by $m = \pm$, denoting the eigenvalues of the mirror operator $M_y$. We label the gap $\Delta_{n}^{m}$ by the band index $n$ and the mirror eigenvalue $m = \pm 1$ of the right-moving state $k_{F_n}>0$.}
\label{fig:spin_split_FS}
\end{figure}
As shown in Fig.~\ref{fig:phasediagram}, the gap closing point at $\phi = \pi$ and $E_{Z,||} =0$ expands into a fan-shaped region in the phase diagram with phase boundaries which
move away from $\phi = \pi$ with slopes which are linearly proportional with $E_{Z,||}$ and decrease with increasing $\mu$. To understand this fan, in the following we derive analytically the dependence of the superconducting gap in a given subband $n$ on $E_{Z,||}$ and $\phi$. For simplicity, we work in the limit where $\Delta, E_{Z,||} \ll \alpha k_F \ll \mu$. The dispersion of the 2DEG, shown in Fig.~\ref{fig:spin_split_FS}(a), exhibits two concentric circular Fermi surfaces. SOC locks the spin orientation to the momentum, such that the outer and inner Fermi surfaces (labeled by $\eta_1 = \pm 1$) have different in-plane spin orientations.
When a Zeeman field is applied, the spin tilts towards the Zeeman field direction. Moreover, to the leading order in $E_{Z,||}/(\alpha k_F)$ the Zeeman field also shifts the two Fermi surfaces uniformly along $k_y$ in the opposite direction by $E_{Z,||}/(\hbar v_{F})$ [see Fig.~\ref{fig:spin_split_FS}(a)].

We now take into account the finite size of the system in the $y$ direction.
We denote the transverse wave functions of the normal Hamiltonian ($\Delta = 0$) by $\varphi^{m}_{n, k_{F_{n}},s} (y)$,
where $n$ is the band index, $s=\uparrow,\downarrow$, and we label each subband according to the
$M_y$ eigenvalue ($m = \pm$) of the state at $k_x = +k_{F_{n}}$ in the limit $E_{Z,||}=0$ [see Fig.~\ref{fig:spin_split_FS}(b)]. A weak Zeeman field mixes the two mirror eigenvalues and opens a gap at $k_x=0$ but does not strongly affect the wave functions at $k_x = k_{F_n}$, such that we may keep using the $\pm$ labeling of the subbands. The walls at $y=\pm(W/2 + W_\text{SC})$ mix states with different values of $\eta_1$ (see Sec.~\ref{sec:rashba} of Ref.~\cite{suppl} for the explicit expression of the wave function).

Proximitizing the 2DEG with SCs induces intraband pairing potentials $\Delta^{\pm}_{n}$ [see Fig.~\ref{fig:spin_split_FS}(b)]; In the limit $W_{\mathrm{SC}} < \xi$, we may neglect the interband matrix elements of the pairing potential.
The pairing potentials $\Delta^\pm_{n}$ can be obtained from the first-order degenerate perturbation theory, and are given by~(see Sec.~\ref{sec:gap} of Ref.~\cite{suppl})
\begin{align}\label{eq:Deltann}
\Delta^\pm_{n} = \frac{1}{W_{\mathrm{SC}}+W/2} \int dy \Delta(y) G_{n}(y) F_{n}^\pm(y),
\end{align}
where
\begin{subequations}
\begin{align}\label{eq:Fny}
G_n(y) &= \sin^2\left[\frac{n\pi (y+ W_{\mathrm{SC}}+W/2)}{W_{\mathrm{SC}}+W/2}\right],\\
F^{\pm}_{n}(y) &= \varphi^{\pm*}_{n,k_{F_{n}},\uparrow}(y) \varphi^{\mp*}_{n,-k_{F_{n}},\downarrow}(y) - (\uparrow \leftrightarrow \downarrow).
\end{align}
\end{subequations}

To the leading order in Zeeman energy, the intraband pairing potential for the $n$th band is
\begin{widetext}
\begin{align}\label{eq:gap}
\Delta_{n}^{\pm} &=\Delta\frac{W_{\mathrm{SC}}}{2W_{\mathrm{SC}}+W}\left[(1+A_n)\cos\left(\frac{\phi}{2}\right)+\frac{E_{Z,||}}{\alpha k_{F_n}}(B_n\pm C_n) \sin\left(\frac{\phi}{2}\right)\right],
\end{align}
\end{widetext}
where $A_n$ is a function of $k_{F_{n}}$, $W$ and $W_{\mathrm{SC}}$, while $B_n$ and $C_n$ are functions of $\alpha$, $v_{F_n}$, $k_{F_{n}}$, $W$, and $W_{\mathrm{SC}}$ [see Eqs.~\eqref{eq:anbncn} and~\eqref{eq:anbncnsimp} in Ref.~\cite{suppl}].
The zeroth-order term of the gap in the Zeeman energy can be understood intuitively as follows. For JJs with narrow SCs ($W_{\mathrm{SC}} \ll \xi$),  electrons undergo multiple normal reflections from the edges of the SCs before they can be Andreev reflected. As a result, the gap is the average of the left and right superconducting gaps, i.e., $\Delta_n^{\pm} \propto \Delta (e^{-i\phi/2}+e^{i\phi/2})/2$ which vanishes at $\phi = \pi$. This gap closing also follows from the fact that the Hamiltonian respects the mirror and time-reversal symmetries at $\phi = \pi$ for $E_{Z,||} = 0$ which implies that $F^{\pm}_{n}(y) = F^{\pm}_n(-y)$~[see Ref.~\cite{suppl} for details]. Since $F^{\pm}_{n}(y)$ and $G_{n}(y)$ are even functions of $y$ while $\Delta(y)$ is an odd function, $\Delta_n^\pm = 0$ at $\phi = \pi$ and $E_{Z,||} = 0$ [see Eq.~\eqref{eq:Deltann}].

Expanding Eq.~\eqref{eq:gap} around $\phi = \pi$, we have the gap-closing points moving away from $\phi = \pi$ by
\begin{align}
\delta \phi_{n}^{\pm} &= \frac{2}{(1+A_n)}\frac{E_{Z,||}}{\alpha k_{F_n}} (B_n \pm C_n).
\end{align}
Thus, inside the fan in the BDI phase diagram, the gap closing lines of each subband move away from $\phi = \pi$ with slopes which are inversely proportional to $k_{F_n}$ (see also Fig.~\ref{fig:phasediagram}). Since these are gap closings at $k_x = \pm k_{F_n}$, they are accompanied by changes in $Q_\mathbb{Z}$ by $\pm 2$, but do not affect $Q_{\mathbb{Z}_2}$. 

As $E_{Z,||}$ increases, the fan of BDI phase boundaries intersects the class D phase boundary where $Q_{\mathbb{Z}_2}$ changes. As seen in Fig.~\ref{fig:phasediagram}, at each of these intersections, either three or four different phases meet.
The four-phase intersection points signify simultaneous gap closings at both $k_x=\pm k_{F_n}$ and $k_x= 0$.
The three-phase intersection points happen when two gap closings at $k_{x} = \pm k_{F_n}$ are moved by varying $E_{Z,||}$ and $\phi$, merge at $k_{x} = 0$, and get lifted  (see Sec.~\ref{sec:phasebound} of Ref.~\cite{suppl} for details).

The BDI symmetry can be broken by applying a transverse in-plane magnetic field (along $y$), disorder that breaks the mirror symmetry, or if the two SCs have different gaps or different widths. Applying a transverse Zeeman field tilts the spectrum, which reduces the gap and results in gapless regions (see Sec.~\ref{sec:transB} of Ref.~\cite{suppl}). On the other hand, the gap-closing points at $k_x = \pm k_{F_n}$ are lifted when the BDI symmetry is broken by disorder~\cite{Haim2018} or an asymmetry of the left and right SCs~\cite{Pientka2017Topological,Hell2017Two,suppl}. Breaking the BDI symmetry also results in the hybridization of MZMs residing at the junction's end, leaving either zero or one MZM at each end (see Sec.~\ref{sec:breakBDI} of Ref.~\cite{suppl}). We note that the \textit{exact} BDI symmetry for planar JJs is preserved as long as the left-right symmetry is not broken, independent of the ratio of $W$ and $W_{\mathrm{SC}}$ to the spin-orbit length  [$\ell_{\mathrm{SO}} = \hbar^2/(m^*\alpha)$]. On the other hand, for the case where the left-right symmetry is broken, there is a transition from a class D to an \textit{approximate} BDI symmetry when $W+W_{\mathrm{SC}}$ drops below $\ell_{\mathrm{SO}}$, similar to the nanowire case~\cite{Tewari2012Topological,Diez2012Andreev}.

\begin{figure}
\centering
{\includegraphics[width = \linewidth]{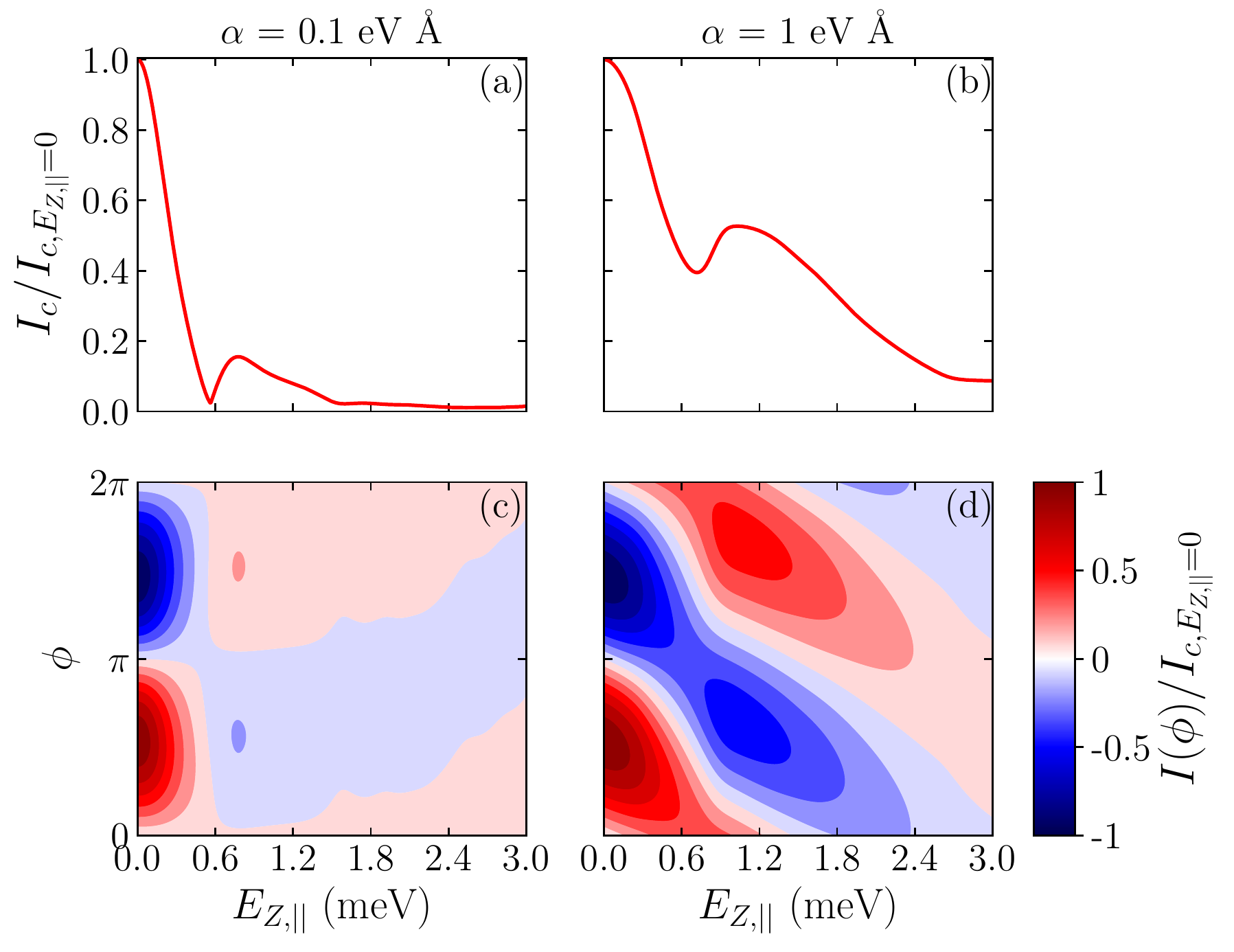}}
\caption{Upper panel: critical current $I_c$ as a function of Zeeman energy $E_{Z,||}$ for different SOC strengths: (a) $\alpha = 0.1$ eV {\AA} and (b) $\alpha = 1$ eV {\AA}. Lower panel: Josephson current as a function of $\phi$ and $E_{Z,||}$ for (c) $\alpha = 0.1$ eV{\AA} and (d) $\alpha = 1$ eV {\AA}. Here, $\mu$ is the same as in Fig.~\ref{fig:phasediagram}(c), and the temperature is $T = 0.3 \Delta/k_\mathrm{B}$. For $\alpha=0.1$ eV {\AA}, $I_c$ exhibits a minimum at $E_{Z,||}^c \approx 0.55$ meV. The minimum of the critical current does not coincide with the class D TPT, which occurs at $E_{Z,\parallel}\approx 0.27$ meV [see Fig.~\ref{fig:phasediagram}(c)].
As $\alpha$ increases, the minimum becomes more shallow [(b)]. For small $\alpha$, there is an abrupt shift by nearly $\pi$ in the current-phase relation $I(\phi)$ at $E_{Z,||}^c$, while for large $\alpha$, $I(\phi)$ has a gradual phase shift with $E_{Z,||}$ [see (b) and (d), respectively].}
\label{fig:Josephson}
\end{figure}

Next, we calculate the Josephson current (see Sec.~\ref{sec:Josephson} of Ref.~\cite{suppl} for details),
\begin{align}\label{eq:Iphi}
I(\phi) = \frac{2e}{\hbar}\frac{d\mathcal{F}}{d\phi} = -\frac{4e}{\hbar}\sum_{j} \tanh\left(\frac{E_{j}}{2k_{\mathrm{B}}T} \right) \frac{d{E_{j}}}{d\phi},
\end{align}
where $\mathcal{F}$ is the free energy of the system, $T$ is the temperature, and $E_j$ are the eigenvalues of the Hamiltonian.
The critical current is
\begin{align}\label{eq:Iceven}
I_c &= \max_\phi I(\phi).
\end{align}
Figure~\ref{fig:Josephson} shows $I_c$ and $I(\phi)$ as a function of $E_{Z,||}$ for a JJ with narrow leads at temperature $T = 0.3\Delta/k_\mathrm{B}$, and for two different values of $\alpha$. The critical current oscillates as a function of $E_{Z,||}$ with an amplitude that decays with $E_{Z,||}$. For small $\alpha$, e.g., $\alpha = 0.1$ eV$\mathrm{\AA}$ [Figs.~\ref{fig:Josephson}(a) and~\ref{fig:Josephson}(c)], at the critical Zeeman field where the critical current exhibits a minimum, the phase at which the free energy is minimal changes from $\phi \approx 0$ to $\phi \approx\pi$.  Unlike JJs with wide SCs~\cite{Pientka2017Topological}, this phase jump does not necessarily imply a TPT due to the weak dependence of the class D TPT on $\phi$ [Fig. \ref{fig:phasediagram}(c)]. For larger values of SOC, e.g., $\alpha = 1$ eV$\mathrm{\AA}$ [Figs.~\ref{fig:Josephson}(b) and~\ref{fig:Josephson}(d)], the critical current exhibits a minimum with a shallower depth and at a larger critical Zeeman field. This minimum, however, is not accompanied by a discontinuous change of $\phi$ that minimizes the free energy. To understand this, we can calculate the energy-phase relation of the junction perturbatively in
$\Delta$, for two different limits: $\alpha k_{F_n} \gg E_{Z,||}$ and $\alpha k_{F_n} \ll E_{Z,||}$ (see Sec.~\ref{sec:Josephson} of Ref.~\cite{suppl}).

In conclusion, we have studied topological superconductivity in planar JJs with narrow SCs and how it crosses over to the nanowire case. As the width of SC leads gets narrower, the strength of normal reflections from the SC edges increases, which renders the class D phase diagram to depend strongly on the chemical potential and more weakly on the superconducting phase difference. On the other hand, the BDI phase diagram is strongly dependent on the superconducting phase difference. Finally, we show that contrary to the wide lead case, the minima in the critical current of JJs with narrow leads do not necessarily indicate TPTs.
These results are directly relevant to recent experiments~\cite{Antonio2018Evidence}, and elucidate the consequences of the BDI symmetry on the phase diagram of these systems.

\begin{acknowledgements}
This work was supported by NSF-DMR-MRSEC 1420709. A.S. and E.B. are supported by CRC 183 of the Deutsche Forschungsgemeinschaft. A.S. acknowledges support from the Israel Science Foundation, the European Research Council (Project LEGOTOP), and Microsoft Station Q.
\end{acknowledgements}

\bibliography{manuscript}


\onecolumngrid
\vspace{1cm}
\begin{center}
{\bf\large Supplemental Material for ``Topological superconductivity in planar Josephson junctions: Narrowing down to the
nanowire limit"}
\end{center}
\vspace{0.5cm}

\setcounter{secnumdepth}{3}
\setcounter{equation}{0}
\setcounter{figure}{0}
\renewcommand{\theequation}{S-\arabic{equation}}
\renewcommand{\thefigure}{S\arabic{figure}}
\renewcommand\figurename{Supplementary Figure}
\renewcommand\tablename{Supplementary Table}
\newcommand\Scite[1]{[S\citealp{#1}]}
\newcommand\Scit[1]{S\citealp{#1}}

\makeatletter \renewcommand\@biblabel[1]{[S#1]} \makeatother

\section{Tight Binding Hamiltonian}\label{sec:tbh}
The Hamiltonian of Eq.~\eqref{eq:HSM} can be written in the tight-binding form as
\begin{equation}
H_{\mathrm{TB}} = H_0 + H_{\mathrm{SOC}} + H_{Z} + H_\Delta,
\end{equation}
where
\begin{align}
H_0 &=  (4t - \mu)\sum_{i = 1}^{L_x}\sum_{j = 1}^{2W_\mathrm{SC}+W} \sum_{s = \uparrow,\downarrow} \psi_{i,j,s}^\dagger \psi_{i,j,s} -t \sum_{\langle i j, i' j'\rangle} \sum_{s = \uparrow,\downarrow} \left[\psi_{i,j,s}^\dagger \psi_{i',j',s} + \mathrm{H.c.}\right], \nonumber\\
H_{\mathrm{SOC}} &= i\alpha_R \sum_{s,s' = \uparrow,\downarrow} \left[\sum^{L_x-1}_{i=1} \sum_{j=1}^{2W_{\mathrm{SC}}+W} \psi_{i+1,j,s}^\dagger \sigma_{y}^{s,s'}\psi_{i,j,s'} - \sum^{L_x}_{i=1} \sum_{j=1}^{2W_{\mathrm{SC}}+W-1} \psi_{i,j+1,s}^\dagger \sigma_{x}^{s,s'}\psi_{i,j,s'} - \mathrm{H.c.}\right], \nonumber\\
H_{Z} &= E_{Z,||}\sum_{i = 1}^{L_x} \sum_{j = 1}^{2W_{\mathrm{SC}}+W} \sum_{s,s'=\uparrow,\downarrow}   \left[\psi^{\dagger}_{i,j,s}\sigma_x^{s,s'}\psi_{i,j,s'}+ \mathrm{H.c.}\right], \nonumber\\
H_{\Delta} &= \Delta e^{-i\phi/2}\sum_{i = 1}^{L_x}\sum_{j=1}^{W_{\mathrm{SC}}}  \psi_{i,j,\uparrow} \psi_{i,j,\downarrow} + \Delta e^{i\phi/2} \sum_{i = 1}^{L_x}\sum_{j=W_{\mathrm{SC}}+W+1}^{2W_{\mathrm{SC}}+W}\psi_{i,j,\uparrow} \psi_{i,j,\downarrow} + \mathrm{H.c.},
\end{align}
with $\psi_{i,j,s}^\dagger$ ($\psi_{i,j,s}$) being the creation (annihilation) operator of an electron with spin $s$ on site $(i,j)$ where $ 1 \leq i \leq L_x$ and $1 \leq j \leq 2W_{\mathrm{SC}}+W$. The hopping strength is denoted by $t$ (for the numerical simulation in this Rapid Communication, we use $t = 14.6$ meV)  where $t = \hbar^2/(2ma^2)$ with $a$ being the lattice constant and the spin-orbit coupling strength is denoted by $\alpha_R$ where $\alpha_R = \alpha/(2a)$. The Zeeman field $E_{Z,||}$ is along the $x$ direction and is taken to be uniform throughout  the system. The proximity-induced superconductivity $\Delta$ is nonzero only for $1\leq j \leq W_{\mathrm{SC}}$ and $W_{\mathrm{SC}}+W+1\leq j \leq 2W_{\mathrm{SC}}+W$.

In the limit where $L_x\rightarrow \infty$, the Hamiltonian can be Fourier transformed using $\psi_{k_x,j,s} = (L_x)^{-1/2}\sum_{i'}e^{ik_xi'a}\psi_{i',j,s}$ as
\begin{align}
H_{\mathrm{TB}} &= \int dk_x [H_0(k_x) + H_{\mathrm{SOC}}(k_x) + H_Z(k_x) + H_\Delta (k_x)],\nonumber\\
H_0(k_x) &=  \left[4 t  - (\mu+ 2t\cos (k_xa))\right]\sum_{j = 1}^{2W_{\mathrm{SC}}+W} \sum_{s = \uparrow,\downarrow}   \psi_{k_x,j,s}^\dagger \psi_{k_x,j,s} -t \sum_{\langle  j,  j'\rangle} \sum_{s = \uparrow,\downarrow} [\psi_{k_x,j,s}^\dagger \psi_{k_x,j',s} + \mathrm{H.c.}],\nonumber\\
H_{\mathrm{SOC}}(k_x) &= \sum_{s,s' = \uparrow,\downarrow} \left[ 2\alpha_R\sin (k_xa)\sum_{j=1}^{2W_{SC}+W}  \psi_{k_x,j,s}^\dagger \sigma_{y}^{s,s'}\psi_{k_x,j,s'} -\left( i\alpha_R \sum_{j=1}^{2W_{SC}+W-1}  \psi_{k_x,j+1,s}^\dagger \sigma_{x}^{s,s'}\psi_{k_x,j,s'} + \mathrm{H.c.}\right)\right], \nonumber\\
H_Z(k_x) &= E_{Z,||}\sum_{s,s' = \uparrow,\downarrow} \sum_{j = 1}^{2W_{SC}+W}  \left(\psi^{\dagger}_{k_x,j,s}\sigma_x^{s,s'}\psi_{k_x,j,s'}+ \mathrm{H.c.}\right), \nonumber\\
H_{\Delta}(k_x) &= \Delta e^{-i\phi/2}\sum_{j=1}^{W_{SC}} \psi_{k_x,j,\uparrow} \psi_{-k_x,j,\downarrow} +\Delta e^{i\phi/2} \sum_{j=W_{SC}+W+1}^{2W_{SC}+W} \psi_{k_x,j,\uparrow} \psi_{-k_x,j,\downarrow} + \mathrm{H.c.}
\end{align}

We use the tight-binding Hamiltonian $H_{\mathrm{TB}}$ to calculate the phase diagram, gap, energy spectrum and Josephson current. The numerical simulations on this tight-binding Hamiltonian are performed using the Kwant package~\cite{groth2014kwant}.

\section{Dependence of class D Phase diagram on the chemical potential and superconducting phase difference}\label{sec:classDphasediagram}

\begin{figure}[h!]
\centering
{\includegraphics[width = 0.9\linewidth]{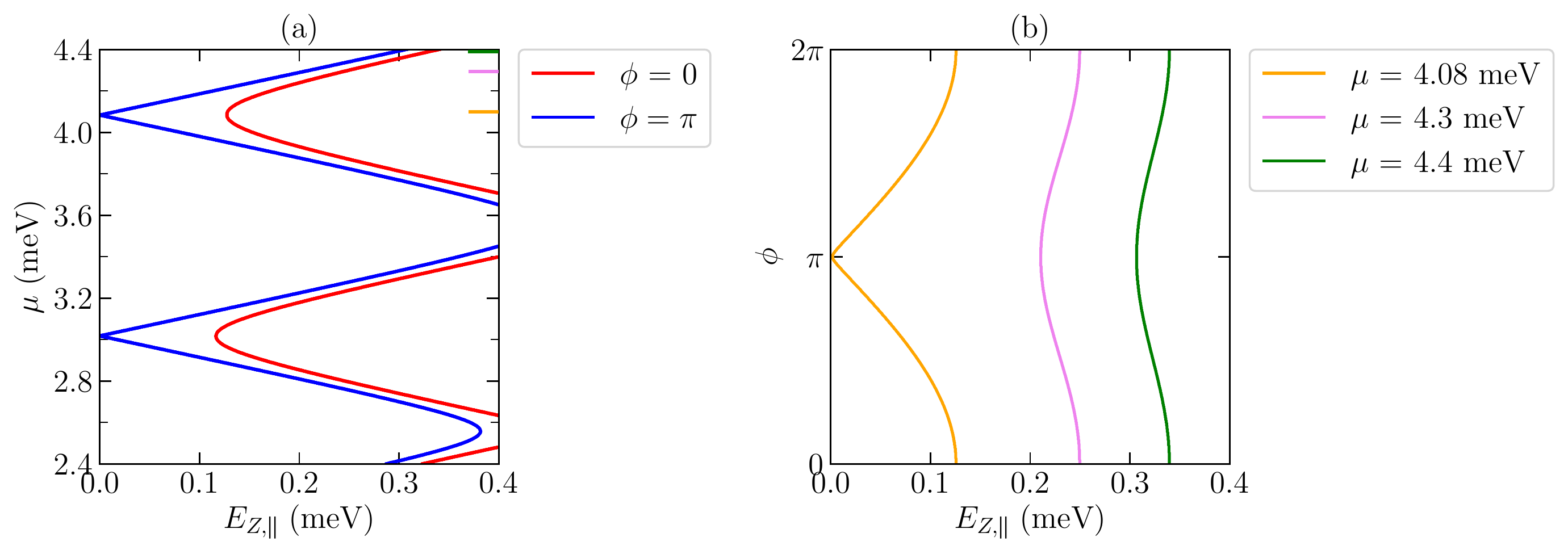}}
\caption{ (a) Class D phase boundary lines as functions of chemical potential $\mu$ and Zeeman field $E_{Z,||}$ for two different values of superconducting phase difference: $\phi = 0$ and $\phi = \pi$. (b) Class D phase boundary lines as functions of superconducting phase difference $\phi$ and Zeeman field $E_{Z,||}$ for different values of chemical potential $\mu$. The phase boundary lines separate the trivial region ($Q_{\mathbb{Z}_2}$ = 1) at lower Zeeman fields from the topological region ($Q_{\mathbb{Z}_2}$ = -1) at higher Zeeman fields. The phase diagrams are calculated by computing the topological invariant ($Q_{\mathbb{Z}_2} = \sgn [\mathrm{Pf}(\mathcal{H}_{k_x=\pi}\sigma_y\tau_y)/\mathrm{Pf}(\mathcal{H}_{k_x=0}\sigma_y\tau_y)]$) following Refs.~\cite{Tewari2012Topological,Kitaev2001Unpaired}. The dependence on the superconducting phase difference $\phi$ [(b)] is stronger for the case where the chemical potential enters a new subband, e.g., $\mu = 3.02$ meV and $\mu = 4.08$ meV [see (a)]. The parameters used are the same as those used in Fig.~\ref{fig:phasediagram} of the main text, i.e., $m^* = 0.026m_e$, $\alpha = 0.1$ eV{\AA}, $\Delta = 0.15$ meV, $W = 80$ nm, and $W_{\mathrm{SC}} = 160$ nm.}
\label{fig:Wlistmucontour}
\end{figure}

Figure~\ref{fig:Wlistmucontour} shows (a) class D boundary lines as functions of Zeeman field $E_{Z,||}$ and chemical potential $\mu$ for two different values of superconducting phase difference: $\phi = 0$ and $\phi = \pi$ and (b) class D boundary lines as functions of Zeeman field $E_{Z,||}$ and superconducting phase difference $\phi$ for several values of chemical potential $\mu$. The class D boundary lines separate the trivial phase at low Zeeman field from the topological phase at high Zeeman field. We note that at $E_{Z,||} = 0$ and $\phi = \pi$, the gap closing points at $k_x = \pm k_{F_{n,x}}$, which signify a BDI phase transition, can also coincide with the gap closing point at $k_x = 0$ which signifies a class D phase transition. This happens when the chemical potential enters a new subband [see the blue lines in Fig.~\ref{fig:Wlistmucontour}(a) for $\mu = 3.02$ meV and $\mu = 4.08$ meV]. We note that the zero critical field for $\phi =\pi$ at these fine-tuned chemical potentials exist only when the BDI symmetry is preserved. Breaking the BDI symmetry will  move the critical field from zero to some finite values~\cite{Antonio2018Evidence}. Since the standard time reversal $T$ is broken for $\phi$ away from $\pi$, the gap opens and increases in magnitude as $\phi$ is tuned away from $\pi$. To close the gap at $\phi$ away from $\pi$, a finite Zeeman field is required. As a result, the class D phase boundary (as a function of $\phi$ and $E_{Z,||}$) has a cusp at $\phi = \pi$ and $E_{Z,||} = 0$ [as shown in Fig.~\ref{fig:Wlistmucontour}(b) for $\mu$ = 4.08 meV]. The dependence of the class D phase diagram on the superconducting phase difference becomes weaker when the chemical potential is tuned away from this special point, i.e., when the band bottom of the occupied subband closest to the chemical potential moves further away from the chemical potential [see Fig.~\ref{fig:Wlistmucontour}].

In general, narrowing the SC leads makes the class D phase boundaries  of a planar JJ depend more on the chemical potential and less  on the superconducting phase difference due to the enhancement of multiple normal reflections from the interface of the superconductors with the vacuum. The narrower is the JJ, the greater is the amplitude of the normal reflections. In the limit where the width is sufficiently narrow ($W_{\mathrm{SC}} \ll \xi$), the phase diagram of JJs with narrow leads will be similiar to that of multiband nanowires~\cite{Lutchyn2011Search,Stanescu2011Majorana}.

\section{Wave function of a Rashba particle in a strip}\label{sec:rashba}
In order to derive the dependence of the superconducting gap on the phase difference and the Zeeman field, we first calculate the wave function of a particle with Rashba SOC and Zeeman field in a strip of width $L_y$ with infinite potential walls. Due the confinement in the $y$ direction, the energy spectrum consists of multiple bands where we label each of them by an index $n$. In the following, we are going to work in the limit where $\Delta \ll E_{Z,||} \ll \alpha k_{F_n} \ll \varepsilon_{F_n}$ and solve for the wave function perturbatively in the Zeeman energy $E_{Z,||}$.

As we consider a system which is translationally invariant along $x$ ($L_x \rightarrow \infty$), we can write the wave function as
\begin{equation}
\Phi_n(x,y) = e^{ik_x x}\varphi_{n,k_x}(y),
\end{equation}
where $\varphi_{n,k_x}(y) = (\varphi_{n,k_{x},\uparrow}(y),\varphi_{n,k_{x},\downarrow}(y))^T$ are two-component spinors of the $n$th band. For the case of zero Zeeman field, we have the Fermi wave vector for the $n$th band as $k_{F_n} = \sqrt{k_{F_{n,x}}^2 + k_{F_{n,y}}^2}$ at $E = \varepsilon_{F_n}$ satisfying
\begin{equation}\label{eq:EFn}
\varepsilon_{F_n} = \frac{\hbar^2(k_{F_n} \mp k_\mathrm{SO})^2}{2m^*},
\end{equation}
where $k_\mathrm{SO} = m^*\alpha/\hbar^2$. The Fermi surface consists of two concentric circles with radius $k_{F_n,\pm} = k_{F_n}\pm k_{\mathrm{SO}}$ as depicted in Fig.~\ref{fig:spin_split_FS}(a) of the main text.
At fixed $k_{F_{n,x}}$, there are four values of $k_{F_{n,y}}$ satisfying Eq.~\eqref{eq:EFn}, which are denoted by $k_{F_{n,y},\eta_{1},\eta_{2}} =\eta_2 k_{F_{n,y},\eta_1}$ where $\eta_{1,2} = \pm$ and
\begin{equation}\label{eq:ky}
k_{F_{n,y},\pm} = \sqrt{(\pm k_\mathrm{SO} + \sqrt{2m^*\varepsilon_{F_n}}/\hbar)^2 - k_{F_{n,x}}^2}.
\end{equation}
The subscripts $\eta_1 = +$ and $\eta_1 = -$ denote the outer and inner Fermi surfaces, respectively, while the subscript $\eta_2 = \pm$ denotes the sign of $k_y$. Due to the spin-orbit momentum locking, the spin rotates along the Fermi surface with the Rashba-induced spin-rotation angle $\theta_{n,\eta_1}$ given by 
\begin{equation}\label{eq:theta}
e^{i\theta_{n,\eta_{1}}} = \eta_1\frac{k_{F_{n,x}} + i  k_{F_{n,y,\eta_{1}}}}{k_{F_n,\eta_1}}.
\end{equation}
In the following, we are going to solve for the spinor $\varphi_{n,k_{F_{n,x}}}(y)$ perturbatively in $E_{Z,||}$ in the regime near $E_{Z,||} = 0$ where the fan in the phase diagram emerges. To the first order in the Zeeman energy, the spinor can be written as
\begin{equation}
\varphi_{n,k_{F_{n,x}}}(y) = \widetilde{\varphi}_{n,k_{F_{n,x}}}(y) + \delta\varphi_{n,k_{F_{n,x}}}(y),
\end{equation}
where $\widetilde{\varphi}_{n,k_{F_{n,x}}}(y)$ is the zeroth-order and $\delta\varphi_{n,k_{F_{n,x}}}(y)$ is the first-order perturbed wave function due to the Zeeman energy.
At $E_{Z,||} = 0$, the Hamiltonian commutes with the mirror operator $M_y = -\sigma_y\times (y\rightarrow -y)$ and we can label the zeroth-order spinor as $\widetilde{\varphi}^m_{n,k_{F_{n,x}}}(y)$ where $m = 1$ and $m = -1$ correspond to the even and odd eigenstates of the mirror operator $M_y$, respectively:
\begin{equation}\label{eq:my}
M_y\widetilde{\varphi}^m_{n,k_{F_{n,x}}}(y) = m \widetilde{\varphi}^m_{n,k_{F_{n,x}}}(-y).
\end{equation}

Due to the confinement in the $y$ direction, the spinor at momentum $k_{F_{n,x}}$ is a superposition of four components, with amplitudes corresponding to the spins at four different $y$ Fermi momenta ($\pm k_{F_{n,y},\pm}$) satisfying Eq.~\eqref{eq:ky}, i.e.,
\begin{equation}\label{eq:wfSOC}
\widetilde{\varphi}^m_{n,k_{F_{n,x}}}(y) \equiv \left(\begin{matrix} \widetilde{\varphi}^m_{n,k_{F_{n,x}},\uparrow}(y)\\ \widetilde{\varphi}^m_{n,k_{F_{n,x}},\downarrow}(y) \end{matrix}\right)= \frac{1}{2}\sum_{\eta_1,\eta_2 = \pm} a^{m}_{n,(\eta_1,\eta_2)} e^{i\eta_2k_{F_{n,y},\eta_1} y} \left(\begin{matrix} e^{-i\eta_2\theta_{n,\eta_1}/2} \\ -i e^{i\eta_2\theta_{n,\eta_1}/2} \end{matrix} \right),
\end{equation}
where $a^{m}_{n,(\eta_1,\eta_2)} (\eta_{1,2} = \pm)$ are coefficients of each $k_{F_{n,y}}$ mode. The coefficients $a^{m}_{n,(\eta_1,\eta_2)}$ are determined from the boundary conditions:
\begin{equation}\label{eq:boundcond}
\widetilde{\varphi}^{m}_{n,k_{F_{n,x}}}(\pm L_y/2) = \left(\begin{matrix} 0 \\ 0 \end{matrix} \right),
\end{equation}
where the walls are at $\pm L_y/2 = \pm (W_{\mathrm{SC}}+ W/2)$. The number of coefficients can be reduced by using the mirror reflection symmetry $M_y = -\sigma_y\times(y\rightarrow -y)$. Since the spinor is either even or odd under reflection
\begin{equation}\label{eq:my}
M_y\widetilde{\varphi}_{n,k_{F_{n,x}}}^{\pm}(y) = \pm  \widetilde{\varphi}_{n,k_{F_{n,x}}}^{\pm}(-y),
\end{equation}
we then have
\begin{equation}\label{eq:reflection}
a^{\pm}_{n,(\eta_1,\eta_2)} = \pm a^{\pm}_{n,(\eta_1,-\eta_2)}.
\end{equation}
The boundary condition [Eq.~\eqref{eq:boundcond}] implies that the spinor [Eq.~\eqref{eq:wfSOC}] for the even mirror eigenstate obeys
\begin{equation}\label{eq:ratioaeven}
\frac{a^{+}_{n,(+,+)}}{a^{+}_{n,(-,+)}} = -\frac{\cos \left(\frac{k_{F_{n,y},-}L_y-\theta_n,{-}}{2} \right)}{\cos \left(\frac{k_{F_{n,y},+}L_y-\theta_{n,+}}{2}\right)} = -\frac{\cos \left(\frac{k_{F_{n,y},-}L_y+\theta_{n,-}}{2} \right)}{\cos \left(\frac{k_{F_{n,y},+}L_y+\theta_{n,+}}{2}\right)}.
\end{equation}
The equation for the odd mirror eigenstate is identical to Eq.~\eqref{eq:ratioaeven} except with cos replaced by sin:
\begin{equation}\label{eq:ratioaodd}
\frac{a^{+}_{n,(+,+)}}{a^{+}_{n,(-,+)}} = -\frac{\sin \left(\frac{k_{F_{n,y},-}L_y-\theta_{n,-}}{2} \right)}{\sin \left(\frac{k_{F_{n,y},+}L_y-\theta_{n,+}}{2}\right)} = -\frac{\sin \left(\frac{k_{F_{n,y},-}L_y+\theta_{n,-}}{2} \right)}{\sin \left(\frac{k_{F_{n,y},+}L_y+\theta_{n,+}}{2}\right)}.
\end{equation}
The above derivation for $E > E_{\mathrm{SO}}$ (where $E_{\mathrm{SO}} = \frac{1}{2}m^2\alpha^2/\hbar^2$ is the SOC energy) follows the derivation for $E < E_{\mathrm{SO}}$ given in the Appendix A of Ref.~\cite{Berg2012Electronic}.

When Zeeman field is introduced, it shifts the center of momentum of the Fermi surfaces and tilts the Rashba-induced spin-rotation angle towards the magnetic field direction [see Fig.~\ref{fig:spin_split_FS}(a) of the main text]. To the first order in $E_{Z,||}$, the change in the center of momentum of the inner ($\eta_1 = -$) and outer ($\eta_1 = +$) Fermi surfaces is given by
\begin{equation}
\delta k_{F_{n,y},\eta_1} = -\eta_1\frac{E_{Z,||}}{\hbar v_{F_{n,\eta_1}}}.
\end{equation}
The Zeeman field rotates the Rashba-induced spin-rotation angle $\theta_{n,\eta_1}$ by
\begin{align}
\delta\theta_{n,\eta_1}(k_x) = \sgn(k_x)\frac{\eta_1 E_{Z,||} \cos\theta_{n,\eta_1}}{\alpha \sqrt{(\kfx)^2+ (k_{F_{n,y},\eta_1})^2}} =  \sgn(k_x)\frac{\eta_1 E_{Z,||}\cos\theta_{n,\eta_1}}{\alpha (\eta_1 k_{\mathrm{SO}}+\sqrt{2m\varepsilon_{F_n}}/\hbar)} &=  \sgn(k_x)\eta_1\frac{E_{Z,||}}{\alpha k_{F_{n,\eta_1}}}\cos\theta_{n,\eta_1}.
\end{align}
To get a simple and compact analytical expression for the wave function, in the following we will focus on the regime where $\alpha k_{F_n} \ll \varepsilon_{F_n}$ where
\begin{subequations}\label{eq:kftheta}
\begin{align}
k_{F_{n,y}} &\equiv k_{F_{n,y},+} = k_{F_{n,y},-}, \\
\delta k_{F_{n,y}} &\equiv \delta k_{F_{n,y},-} = -\delta k_{F_{n,y},+} = \frac{E_{Z,||}}{\hbar v_{F_n}}, \\
\theta_n &\equiv \theta_{n,+}(k_{F_{n,x}}) = \theta_{n,-}(k_{F_{n,x}})-\pi, \\
\delta\theta_n &\equiv \delta \theta_{n,+}(k_{F_{n,x}}) = - \delta \theta_{n,-}(k_{F_{n,x}}) = \frac{E_{Z,||}}{\alpha k_{F_n}}\cos\theta_n.
\end{align}
\end{subequations}
To the first order in Zeeman energy, the change in the spinor can be written as
\begin{align}\label{eq:varphizeeman}
\delta \varphi^m_{n,k_{F_{n,x}}}(y) &= \frac{1}{2}\sum_{\eta_1,\eta_2 = \pm} e^{i\eta_2k_{F_{n,y}}y}\left[\left(\delta a^{m}_{n,(\eta_1,\eta_2)}-i\eta_1 a^{m}_{n,(\eta_1,\eta_2)} \delta k_{F_{n,y}}y\right) \left(\begin{matrix} e^{-i(\eta_2\theta_{n,\eta_1})/2}\\ -i  e^{i (\eta_2\theta_{n,\eta_1} )/2} \end{matrix}\right) -i\eta_1 a^{m}_{n,(\eta_1,\eta_2)} \frac{\delta \theta_n}{2}\left(\begin{matrix} e^{-i(\eta_2\theta_{n,\eta_1})/2}\\ i  e^{i (\eta_2\theta_{n,\eta_1} )/2} \end{matrix}\right) \right],
\end{align}
where $\delta a_{n,(\eta_1,\eta_2)}^{m}$ is the first-order correction to the coefficients $a_{n,(\eta_1,\eta_2)}^{m}$ due to Zeeman energy.

To get the explicit form of $\delta a^m_{n,(\eta_1,\eta_2)}$, we impose the boundary condition on Eq.~\eqref{eq:varphizeeman}:
\begin{equation}\label{eq:boundperturb}
\delta\varphi^m_{n,k_{F_{n,x}}} (\pm L_y/2) = \left(\begin{matrix} 0 \\ 0  \end{matrix}\right).
\end{equation}
Solving for Eq.~\eqref{eq:boundperturb} and using the notation in Eq.~\eqref{eq:kftheta}, we then have
\begin{align}\label{eq:deltaa2}
\delta a^{m}_{n,(++)} &= \frac{a^{m}_{n,(++)} \delta \kfl \cos(k_{F_{n,y}}L_y) + a^{m}_{n,(+-)}\delta k_{F_{n,y}} L_y \cos \theta_n -  a^{m}_{n,(-+)}\delta \theta_n \sin (k_{F_{n,y}}L_y) + a^{m}_{n,(--)} \dkfla \sin \theta_n }{2 \sin(\kfl) },\nonumber\\
\delta a^{m}_{n,(+-)} &= \frac{-a^{m}_{n,(+-)} \delta \kfl \cos(k_{F_{n,y}}L_y) - a^{m}_{n,(++)}\delta k_{F_{n,y}} L_y \cos \theta_n + a^{m}_{n,(--)}\delta \theta_n \sin (k_{F_{n,y}}L_y) -a^{m}_{n,(-+)} \dkfla \sin \theta_n }{2 \sin(\kfl) }, \nonumber\\
\delta a_{n,(-+)}^{m} &= \frac{-a^{m}_{n,(-+)} \delta \kfl \cos(k_{F_{n,y}}L_y) + a^{m}_{n,(--)}\delta k_{F_{n,y}} L_y \cos \theta_n -  a^{m}_{n,(++)}\delta \theta_n \sin (k_{F_{n,y}}L_y) -  a^{m}_{n,(+-)} \dkfla \sin \theta_n }{2 \sin(\kfl) },\nonumber\\
\delta a^{m}_{n,(--)} &=  \frac{a^{m}_{n,(--)} \delta \kfl \cos(k_{F_{n,y}}L_y) - a^{m}_{n,(-+)}\delta k_{F_{n,y}} L_y \cos \theta_n +  a^{m}_{n,(+-)}\delta \theta_n \sin (k_{F_{n,y}}L_y) +  a^{m}_{n,(++)} \dkfla \sin \theta_n }{2 \sin(\kfl)}.
\end{align}
Substituting Eq.~\eqref{eq:reflection} into Eq.~\eqref{eq:deltaa2}, we have
\begin{equation}\label{eq:deltaarel}
\delta a_{n,(\eta_1,-\eta_2)}^\pm(k_x) = \mp \delta a_{n,(\eta_1,\eta_2)}^{\pm}(k_x).
\end{equation}

From Eqs.~\eqref{eq:reflection},~\eqref{eq:ratioaeven}, and~\eqref{eq:ratioaodd}, we have for $k_x = k_{F_{n,x}}$:
\begin{subequations}\label{eq:aeq}
\begin{align}
a^{+}_{n,(++)}(k_{F_{n,x}}) &= a^{+}_{n,(+-)}(k_{F_{n,x}}) =  \sinkm,\\
a^{+}_{n,(-+)}(k_{F_{n,x}}) &= a^{+}_{n,(--)}(k_{F_{n,x}}) = -\coskm,\\
a^{-}_{n,(++)}(k_{F_{n,x}}) &= -a^{-}_{n,(+-)}(k_{F_{n,x}}) =  i\coskp,\\
a^{-}_{n,(-+)}(k_{F_{n,x}}) &= -a^{-}_{n,(--)}(k_{F_{n,x}}) = -i\sinkp.
\end{align}
\end{subequations}
Substituting Eq.~\eqref{eq:aeq} into Eq.~\eqref{eq:deltaa2}, we obtain for $k_x = k_{F_{n,x}}$:
\begin{align}
\delta a^{+}_{n,(++)}(k_{F_{n,x}})  &= \sinkm\delta k_{F_{n,y}}[\cos(\kfl) +\cos\theta_n]+\coskm[\delta\theta_n\sin(\kfl)-\dkfla\sin\theta_n],\nonumber\\
\delta a^{+}_{n,(-+)}(k_{F_{n,x}})  &= \coskm\delta k_{F_{n,y}}[\cos(\kfl)-\cos\theta_n]-\sinkm[\delta\theta_n\sin(\kfl)+\dkfla\sin\theta_n],\nonumber\\
\delta a^{-}_{n,(++)}(k_{F_{n,x}})  &= i\coskp\delta k_{F_{n,y}}[\cos(\kfl) - \cos\theta_n] +i\sinkp[\delta\theta_n\sin(\kfl)+\dkfla\sin\theta_n],\nonumber\\
\delta a^{-}_{n,(-+)}(k_{F_{n,x}})  &= i\sinkp\delta k_{F_{n,y}}[\cos(\kfl)+\cos\theta_n] -i \coskp[\delta\theta_n\sin(\kfl)-\dkfla\sin\theta_n],\nonumber\\
\end{align}
and we use Eq.~\eqref{eq:deltaarel} to obtain the remaining four coefficients.

To obtain $a^{m}_{n,(\eta_1,\eta_2)}(-k_{F_{n,x}})$ and $\delta a^{m}_{n,(\eta_1,\eta_2)}(-k_{F_{n,x}})$, let us derive the relation between these coefficients with their counterparts $a^{m}_{n,(\eta_1,\eta_2)}(k_{F_{n,x}})$ and $\delta a^{m}_{n,(\eta_1,\eta_2)}(k_{F_{n,x}})$ at $k_x = k_{F_{n,x}}$ by using the time-reversal and mirror symmetries.

For $E_{Z,||} = 0$, the Hamiltonian commutes with the standard time-reversal operator $T = -i \sigma_y K$ and the mirror operator $M_y = -\sigma_y\times (y\rightarrow -y)$. Under the standard time-reversal symmetry $T$, the spinor transforms as
\begin{equation}\label{eq:T}
T\widetilde{\varphi}_{n,k_{F_{n,x}}}^{\pm}(y) = e^{i\lambda}\widetilde{\varphi}_{n,-k_{F_{n,x}}}^\pm(y).
\end{equation}
We will fix the gauge to be $\lambda = 0$ so that the superconducting gap is real (see Sec.~\ref{sec:gap} for a proof). Since $\{M_y,T\} = 0$,  we then have
\begin{align}\label{eq:Ttilde}
M_y\left(T \varphi_{n,k_{F_{n,x}}}^{\pm} (y) \right) &= -T\left(M_y \widetilde{\varphi}_{n,k_{F_{n,x}}}^{\pm}(y)\right)\nonumber\\
&=-T\left(M_y \varphi_{n,k_{F_{n,x}}}^{\pm}(y)\right)\nonumber\\
&= \mp \left(T\varphi^{\pm}_{n,k_{F_{n,x}}}(-y)\right).
\end{align}
Eqs.~\eqref{eq:my},~\eqref{eq:T}, and~\eqref{eq:Ttilde} imply that
\begin{subequations}\label{eq:wavefuncT}
\begin{align}
\varphi_{n,-k_{F_{n,x}}}^{\mp}(y) &= T\varphi_{n,k_{F_{n,x}}}^{\pm}(y),\\
\left(\begin{matrix}
\widetilde{\varphi}_{n,-k_{F_{n,x}},\uparrow}^{\mp}(y)\\
\widetilde{\varphi}_{n,-k_{F_{n,x}},\downarrow}^{\mp}(y)
\end{matrix}  \right) &= \left(
\begin{matrix}
-\widetilde{\varphi}^{\pm*}_{n,k_{F_{n,x}},\downarrow}(y) \\
\widetilde{\varphi}^{\pm*}_{n,k_{F_{n,x}},\uparrow}(y)
\end{matrix}
\right),
\end{align}
\end{subequations}
which in turn gives the following relations
\begin{subequations}\label{eq:aT}
\begin{align}
a^{\mp}_{n,(\eta_1,\eta_2)}(-k_{F_{n,x}}) &= \eta_2[a^{\pm}_{n,(\eta_1,-\eta_2)}(k_{F_{n,x}})]^*,\label{eq:aT1}\\
\theta_{n,\eta_1}(-k_{F_{n,x}}) &= \pi- \theta_{n,\eta_1}(k_{F_{n,x}}).\label{eq:thetaT1}
\end{align}
\end{subequations}
Acting upon by the ``effective" time-reversal operator $\widetilde{T} = M_y T = iK\times (y \rightarrow -y)$, the spinor transforms as
\begin{align}\label{eq:effT}
\widetilde{T} \varphi_{n,k_{F_{n,x}}}^{\pm}(y) &= M_y T \varphi_{n,k_{F_{n,x}}}^{\pm}(y) \nonumber\\
&=M_y\varphi_{n,-k_{F_{n,x}}}^{\mp}(y)\nonumber\\
&=\mp\varphi_{n,-k_{F_{n,x}}}^{\mp}(-y).
\end{align}
As a result, we have
\begin{equation}\label{eq:wavefunceffT}
\left(\begin{matrix}
\varphi_{n,-\kfx,\uparrow}^{\mp}(y) \\
\varphi_{n,-\kfx,\downarrow}^{\mp}(y)
\end{matrix} \right) = \mp i\left(\begin{matrix}
\varphi_{n,\kfx,\uparrow}^{\pm*}(-y) \\
\varphi_{n,\kfx,\downarrow}^{\pm*}(-y)
\end{matrix} \right),
\end{equation}
which gives
\begin{subequations}\label{eq:aeffT}
\begin{align}
a^{\mp}_{n,(\eta_1,\eta_2)}(-k_{F_{n,x}}) &= \pm \eta_2[a^{\pm}_{n,(\eta_1,\eta_2)}(k_{F_{n,x}})]^*,\label{eq:aeffT1}\\
\theta_{n,\eta_1}(-k_{F_{n,x}}) &= \pi- \theta_{n,\eta_1}(k_{F_{n,x}}),\label{eq:thetaeffT1}\\
\delta\theta_{n,\eta_1}(-k_{F_{n,x}}) &= - \delta\theta_{n,\eta_1}(k_{F_{n,x}}),\label{eq:dthetaeffT1}\\
\delta a^{\mp}_{n,(\eta_1,\eta_2)}(-k_{F_{n,x}}) &= \pm \eta_2[\delta a^{\pm}_{n,(\eta_1,\eta_2)}(k_{F_{n,x}})]^*.\label{eq:daeffT1}
\end{align}
\end{subequations}

Using the above equations for $a^{m}_{n,(\eta_1,\eta_2)}$ and $\delta a^{m}_{n,(\eta_1,\eta_2)}$, we can then evaluate the spinor to the first order in Zeeman energy as
\begin{equation}\label{eq:varphi}
\varphi^m_{n,k_{F_{n,x}}}(y) = \widetilde{\varphi}^m_{n,k_{F_{n,x}}}(y) + \delta \varphi^m_{n,k_{F_{n,x}}}(y),
\end{equation}
where $\widetilde{\varphi}^m_{n,k_{F_{n,x}}}(y)$ and $\delta\varphi^m_{n,k_{F_{n,x}}}(y)$ are the zeroth order [Eq.~\eqref{eq:wfSOC}] and first-order term [Eq.~\eqref{eq:varphizeeman}] in Zeeman energy, respectively. We are now in the position to derive the dependence of the superconducting gap on the Zeeman field.

\section{Dependence of Superconducting Gap on The Superconducting Phase Difference and Zeeman Field}\label{sec:gap}

In this section, we are going to use the wave function of a Rashba particle obtained in the previous section to derive the dependence of the superconducting gap on the superconducting phase difference and Zeeman field. We begin by writing down the superconductivity term as
\begin{equation}\label{eq:deltay}
H_{\Delta} = \int dy \Delta(y) \left[\psi^\dagger_{k_{F_x},\uparrow}(y) \psi^\dagger_{-k_{F_{x}},\downarrow}(y) - \psi^\dagger_{k_{F_{x}},\downarrow}(y) \psi^\dagger_{-k_{F_x},\uparrow}(y)\right].
\end{equation}
Note that only pairing between modes with opposite momenta at the Fermi energy contributes to the gap. We now incorporate the size quantization along the $y$ direction to take into account the narrow width of the superconducting leads.
We can write down the field operator as
\begin{align}\label{eq:psiy}
\psi_{k_{F_x},\uparrow}(y) &= \sqrt{\frac{2}{L_y}}\sum_{n}^N\sum_{m = \pm} \sin\left(\frac{n\pi(y+L_y/2)}{L_y}\right)\varphi^{m}_{n,k_{F_{n,x}},\uparrow}(y) c_{n,k_{F_{n,x}}}, \nonumber\\
\psi^\dagger_{-k_{F_x},\downarrow}(y) &= \sqrt{\frac{2}{L_y}}\sum_{n}^N\sum_{m = \pm}\sin\left(\frac{n\pi(y+L_y/2)}{L_y}\right) [\varphi^{m}_{n,-k_{F_{n,x}},\downarrow}(y)]^* c_{n,-k_{F_{n,x}}}^\dagger,
\end{align}
where $c_{n,k_{F_{n,x}}}$ is the annihilation operator for an electron in the $n$th band at the Fermi level with momentum $k_{F_{n,x}}$ and $\varphi^{m}_{n,k_{F_{n,x}},s}(y)$ is the spinor for the $m = \pm$ mirror eigenstate of the $n$th band with momentum $k_{F_{n,x}}$ and spin $s = \uparrow/\downarrow$. Since we are interested only in the low-energy modes, only the contribution from those bands which cross the Fermi energy ($n = 1 \cdots N$) are taken into account.

Substituting Eq.~\eqref{eq:psiy} into Eq.~\eqref{eq:deltay} and by noting that only the pairing between modes $\varphi^{\pm}_{n,k_{F_{n,x}},\uparrow/\downarrow}(y)$ and its time-reversed partner $\varphi^{\mp}_{n',-k_{F_{n,x}},\downarrow/\uparrow}(y)$ [Eq.~\eqref{eq:wavefuncT}] will open a gap at the Fermi energy, we have the superconducting term as
\begin{align}\label{eq:Hs}
H_{\Delta} &= \frac{2}{L_y} \sum_{n}\sum_{n'}\sum_{m = \pm}\int dy \Delta(y) G_{nn'}(y) \left[\varphi^{m*}_{n,k_{F_{n,x}},\uparrow}(y) \varphi^{-m*}_{n',-k_{F_{n',x}},\downarrow}(y) - \varphi^{m*}_{n,k_{F_{n,x}},\downarrow}(y) \varphi^{-m*}_{n',-k_{F_{n',x}},\uparrow}(y)  \right]  c^{\dagger,m}_{n,k_{F_{n,x}}} c^{\dagger,-m}_{n',-k_{F_{n',x}}}\nonumber\\
& = \sum_{n}\sum_{n'}\sum_{m = \pm} \Delta^m_{n,n'} c^{\dagger,m}_{n,k_{F_{n,x}}} c^{\dagger,-m}_{n',-k_{F_{n',x}}},
\end{align}
where
\begin{align}\label{eq:Deltan}
\Delta^{\pm}_{nn'} &= \frac{2}{L_y}\int dy  \Delta(y)G_{nn'}(y) F_{nn'}^{\pm}(y),
\end{align}
with
\begin{align}
\Delta(y) &= \Delta e^{i\sgn(y)\phi/2}\Theta(W_{\mathrm{SC}}+W/2-|y|)\Theta(|y|-W/2),\nonumber\\
G_{nn'}(y) &=  \sin\left[\frac{n\pi (y+ L_y/2)}{L_y}\right]\sin\left[\frac{n'\pi (y+ L_y/2)}{L_y}\right],\nonumber\\
 F_{nn'}^{\pm}(y) &=\varphi^{\pm*}_{n,k_{F_{n,x}},\uparrow}(y) \varphi^{\mp*}_{n',-k_{F_{n',x}},\downarrow}(y) - \varphi^{\pm*}_{n,k_{F_{n,x}},\downarrow}(y) \varphi^{\mp*}_{n',-k_{F_{n',x}},\uparrow}(y).
\end{align}

In the following, we are going to work in the limit $W_{\mathrm{SC}} \ll \xi$ where the interband spacing $\hbar v_F/(2W_{\mathrm{SC}}+W)$ is large such that the interband pairing ($\Delta_{nn'}$ for $n \neq n'$) is small and thus can be neglected. Moreover, we will work in the limit where $\Delta \ll E_{Z,||} \ll \alpha k_{F_n} \ll \varepsilon_{F_n}$ and focus on the region where $E_{Z,||}$ is near 0 and $\phi$ is near $\pi$ where the fan in the BDI phase diagram emerges. In this limit we have the relation between the spinors given by Eqs.~\eqref{eq:wavefunceffT} and~\eqref{eq:varphi}, and thus we can write down the intraband pairing ($\Delta_{n} \equiv \Delta_{n,n}$) at $k_{F_{n,x}}$ as
\begin{align}
\Delta_{n,k_{F_{n,x}}}^{\pm} &= \frac{2}{L_y}\int_{-L_y/2}^{L_y/2} dy  \Delta(y) G_{nn}(y) \left[\varphi^{\pm*}_{n,k_{F_{n,x}},\uparrow}(y) \varphi^{\mp*}_{n,-k_{F_{n,x}},\downarrow}(y) - \varphi^{\pm*}_{n,k_{F_{n,x}},\downarrow}(y) \varphi^{\mp*}_{n,-k_{F_{n,x}},\uparrow}(y) \right] \nonumber\\
&=\frac{2}{L_y}\int_{-L_y/2}^{L_y/2} \Delta(y) G_{nn}(y) (\mp i)[\varphi_{n,k_{F_{n,x}},\uparrow}^{\pm*}(y)\varphi_{n,k_{F_{n,x}},\downarrow}^{\pm}(-y) - \varphi_{n,k_{F_{n,x}},\downarrow}^{\pm*}(y)\varphi_{n,k_{F_{n,x}},\uparrow}^{\pm}(-y)]\nonumber\\
&\equiv\frac{2}{L_y}\int_{-L_y/2}^{L_y/2} \Delta(y) G_{nn}(y) F_{n,{k_{F_{n,x}}}}^{\pm}(y).
\end{align}
Since $F_{n,k_{F_{n,x}}}^{\pm}(y) = [F_{n,k_{F_{n,x}}}^{\pm}(-y)]^*$ and $\Delta(y)G_{nn}(y) = [\Delta(-y)G_{nn}(-y)]^*$, we then have
\begin{equation}
\Delta_{n,k_{F_{n,x}}}^{\pm*} = \int dy [\Delta(y)G_{nn}(y)]^* [F_{n,k_{F_{n,x}}}^{\pm}(y)]^* = \int dy \Delta(-y)G_{nn}(-y) F_{n,k_{F_{n,x}}}^{\pm}(-y) = \Delta_{n,k_{F_{n,x}}}^{\pm}.
\end{equation}
So, in the gauge chosen [Eq.~\eqref{eq:wavefunceffT}], the pairing potential is always real. Moreover, since
\begin{align}
\Delta^{m}_{n,k_{F_{n,x}}} c_{n,k_{F_{n,x}}}^{m\dagger} c_{n,-k_{F_{n,x}}}^{-m\dagger} = \Delta^{m}_{n,-k_{F_{n,x}}} c_{n,-k_{F_{n,x}}}^{-m\dagger} c_{n,k_{F_{n,x}}}^{m\dagger},
\end{align}
we have
\begin{align}
\Delta^m_{n,-k_{F_{n,x}}} &= -\Delta^m_{n,k_{F_{n,x}}}, \nonumber\\
F^m_{n,-k_{F_{n,x}}} &= -F^m_{n,k_{F_{n,x}}}.
\end{align}
In the following, we are going to calculate the dependence of the pairing potential $\Delta^\pm_n \equiv \Delta^\pm_{n,k_{F_{n,x}}}$ on the Zeeman energy $E_{Z,||}$ and superconducting phase difference $\phi$. To this end, we expand the spinor to the leading order in the Zeeman field as $\varphi_{n,k_{F_{n,x}},s}^m = \widetilde{\varphi}_{n,k_{F_{n,x}},s}^m + \delta\varphi_{n,k_{F_{n,x}},s}^m$ where we have the pairing correlation $F_{n}^{\pm}(y)\equiv F_{n,k_{F_{n,x}}}^{\pm}(y)$ as
\begin{align}
F_{n}^{\pm}(y) = \widetilde{F}^{\pm}_{n}(y) + \delta F^{\pm}_{n}(y)
\end{align}
with the zeroth- and first-order terms in $E_{Z,||}$ being
\begin{align}
\widetilde{F}^{\pm}_{n}(y) &\equiv \widetilde{\varphi}^{\pm*}_{n,k_{F_{n,x}},\uparrow}(y) \widetilde{\varphi}^{\mp*}_{n,-k_{F_{n,x}},\downarrow}(y) - \widetilde{\varphi}^{\pm*}_{n,k_{F_{n,x}},\downarrow}(y) \widetilde{\varphi}^{\mp*}_{n,-k_{F_{n,x}},\uparrow}(y)\nonumber\\
&= |\widetilde{\varphi}^{\pm}_{n,k_{F_{n,x}},\uparrow}(y)|^2 + |\widetilde{\varphi}^{\pm}_{n,k_{F_{n,x}},\downarrow}(y)|^2,\nonumber\\
\delta F^{\pm}_{n}(y) &= \delta \varphi^{\pm*}_{n,k_{F_{n,x}},\uparrow}(y) \widetilde{\varphi}^{\mp*}_{n,-k_{F_{n,x}},\downarrow}(y) + \delta\varphi^{\mp*}_{n,-k_{F_{n,x}},\downarrow}(y) \widetilde{\varphi}^{\pm*}_{n,k_{F_{n,x}},\uparrow}(y)  - (\uparrow \leftrightarrow \downarrow)\nonumber\\
&= \delta \varphi^{\pm*}_{n,k_{F_{n,x}},\uparrow}(y) \widetilde{\varphi}^{\pm}_{n,k_{F_{n,x}},\uparrow}(y) + \delta \varphi^{\mp*}_{n,-k_{F_{n,x}},\downarrow}(y) \widetilde{\varphi}^{\mp}_{n,-k_{F_{n,x}},\downarrow}(y) + (\uparrow \leftrightarrow \downarrow).
\end{align}

From Eq.~\eqref{eq:wfSOC}, we can calculate the zeroth-order term of the pairing correlation as
\begin{align}\label{eq:Fzero}
\widetilde{F}^{\pm}_{n}(y)&=\frac{1}{4}\left[\left|\sum_{\eta_1,\eta_2=\pm} a^{\pm}_{n,(\eta_1,\eta_2)}e^{i\eta_2(k_{F_{n,y}} y - \theta_{n,\eta_1}/2)}\right|^2 + \left|\sum_{\eta_1,\eta_2=\pm} a^{\pm}_{n,(\eta_1,\eta_2)}e^{i\eta_2(k_{F_{n,y}} y + \theta_{n,\eta_1}/2)}\right|^2 \right]\nonumber\\
&= 1 - \cos(2 k_{F_{n,y}} y)\cos(k_{F_{n,y}} L_y).
\end{align}
Using Eq.~\eqref{eq:Fzero}, we can then evaluate the zeroth order term of the superconducting gap to be
\begin{align}
\widetilde{\Delta}_{n}^{\pm} &=\frac{2}{L_y}\int_{-L_y/2}^{L_y/2} dy \Delta(y)  \sin^2\left[\frac{n\pi(y+L_y/2)}{L_y} \right]  \widetilde{F}^{\pm}_{n,k_{F_{n,x}}}(y) \nonumber\\
&=\Delta\frac{W_{\mathrm{SC}}}{2W_{\mathrm{SC}}+W}\cos\left(\frac{\phi}{2}\right)\nonumber\\
&\hspace{0.5 cm}\times\left\{1+(-1)^n \frac{2W_{\mathrm{SC}}+W}{2 n \pi W_{\mathrm{SC}}}\sin\left(\frac{n\pi W}{ 2W_{\mathrm{SC}}+W} \right)\right.\nonumber\\
&\hspace{1 cm}\left.+\cos[\kfy (2W_{\mathrm{SC}}+W)]\left[-f_0(2\kfy) + \frac{(-1)^n}{2}f_0\left(2\kfy+\frac{2n\pi}{2W_{\mathrm{SC}}+W}\right)+ \frac{(-1)^n}{2}f_0\left(2\kfy-\frac{2n\pi}{2W_{\mathrm{SC}}+W}\right) \right]\right\}
\color[rgb]{0,0,0},
\end{align}
where
\begin{align}
f_0(k_y) &=\frac{1}{W_{\mathrm{SC}}} \int^{W/2+W_{\mathrm{SC}}}_{W/2}dy \cos(k_y y)\nonumber\\
&= \frac{1}{k_y W_{\mathrm{SC}}} \left\{\sin\left[k_y\left(W_{\mathrm{SC}}+\frac{W}{2}\right) \right] - \sin \left[k_y\left(\frac{W}{2}\right)\right] \right\}\nonumber\\
&=  \cos\left[k_y\left(\frac{W_{\mathrm{SC}}+W}{2}\right)\right]\sinc\left[k_y\left(\frac{W_{\mathrm{SC}}}{2}\right)\right].
\end{align}

For the case where $n \gg 1$, we have
\begin{equation}
\widetilde{\Delta}_{n}^{\pm} = \Delta\frac{W_{\mathrm{SC}}}{2 W_{\mathrm{SC}}+W}(1 + A_n)\cos\left(\frac{\phi}{2}\right) ,
\end{equation}
where
\begin{equation}
A_n = - \cos[k_{F_{n,y}}(2W_{\mathrm{SC}}+W)]\cos\left[k_{F_{n,y}}\left(W_{\mathrm{SC}}+W\right)\right]\sinc\left(k_{F_{n,y}} W_{\mathrm{SC}}\right).
\end{equation}

We now calculate the dependence of the pairing correlation on the Zeeman energy. After a lengthy algebra, we obtain the first-order correction to the pairing correlation due to the Zeeman field as
\begin{align}
\delta F^{\pm}_{n}(y) &= 2i\delta k_{F_{n,y}}\left[ L_y \sin(k_{F_{n,y}} L_y)\sin\theta_n\sin(2k_{F_{n,y}} y) \pm 2 \sin\left(\frac{k_{F_{n,y}} L_y}{2}\right) \sin\left(\frac{k_{F_{n,y}} L_y}{2} \mp\theta_n\right)y \cos(k_{F_{n,y}} y)\right]\nonumber\\
&\hspace{2 cm}- 4i\delta\theta_n \left[ \cos\left(\frac{k_{F_{n,y}} L_y}{2}\right)\sin\left(\frac{k_{F_{n,y}} L_y}{2} \mp \theta_n\right)\sin(k_{F_{n,y}} y)\pm\sin(k_{F_{n,y}} L_y)\sin\theta_n \sin(2k_{F_{n,y}} y)  \right].
\end{align}
The first-order correction to the gap due to the Zeeman field is then given by
\begin{align}\label{eq:gapeven}
\delta \Delta_{n}^{\pm} & = \frac{W_{\mathrm{SC}}}{2 W_{\mathrm{SC}}+W} \int_{-L_y/2}^{L_y/2} dy \Delta(y) \sin^2\left(\frac{n\pi(y+L_y/2)}{L_y}\right) \delta F^{\pm}_{n}(y)\nonumber\\
&= \Delta\frac{W_{\mathrm{SC}}}{2 W_{\mathrm{SC}}+W}\sin\left(\frac{\phi}{2}\right)\nonumber\\
&\hspace{0.5 cm}\times\left[ 4(-\dkfl \pm \delta \theta_n) \sin(\kfy L_y)\sin\theta_n \left\{f_2(2\kfy)- \frac{(-1)^n}{2}f_2\left(2\kfy + \frac{2n\pi}{L_y}\right) + \frac{(-1)^n}{2} f_2\left(2\kfy - \frac{2n\pi}{L_y}\right) \right\} \right. \nonumber\\
&\hspace{1 cm}  \mp 8\delta\kfy \sin\left(\frac{\kfy L_y}{2}\right) \sin\left(\frac{\kfy L_y}{2} \mp\theta_n\right) \left\{f_1(\kfy)- \frac{(-1)^n}{2}f_1\left(\kfy + \frac{2n\pi}{L_y}\right) + \frac{(-1)^n}{2} f_1\left(\kfy - \frac{2n\pi}{L_y}\right)\right\} \nonumber\\
&\left.\hspace{1 cm}+ 4 \delta\theta_n  \cos\left(\frac{\kfy L_y}{2}\right)\sin\left(\frac{\kfy L_y}{2} \mp \theta_n\right) \left\{f_2(\kfy)- \frac{(-1)^n}{2}f_2\left(\kfy + \frac{2n\pi}{L_y}\right) + \frac{(-1)^n}{2} f_2\left(\kfy - \frac{2n\pi}{L_y}\right) \right\} \right],
\end{align}
where
\begin{align}
f_1(k_y) &= \frac{1}{W_{\mathrm{SC}}}\int^{W/2+W_{\mathrm{SC}}}_{W/2}dy y\cos(k_y y)\nonumber\\
&= \frac{1}{\wsc}\left[\frac{y\sin(k_y y)}{k_y}\right]^{W/2+W_{\mathrm{SC}}}_{W/2} - \frac{1}{\wsc}\int_{W/2}^{W/2+W_{\mathrm{SC}}} \frac{\sin (k_y y)}{k_y}dy,\nonumber\\
&=\frac{1}{k_y\wsc}\left\{\left(\frac{W}{2}+W_{\mathrm{SC}}\right)\sin\left[k_y\left(\frac{W}{2}+W_{\mathrm{SC}}\right)\right] - \frac{W}{2}\sin\left(k_y\frac{W}{2}\right)\right\} - \frac{2}{k_y^2\wsc}\sin\left[k_y\left(\frac{W+W_{\mathrm{SC}}}{2}\right) \right]\sin\left(k_y\frac{W_{\mathrm{SC}}}{2}\right)\nonumber\\
&=\frac{W}{2}\cos\left[k_y\left(\frac{W_{\mathrm{SC}}+W}{2}\right)  \right]\sinc\left[k_y\left(\frac{W_{\mathrm{SC}}}{2}\right)\right]+\frac{1}{k_y}\left\{\sin\left[k_y\left(\frac{W}{2}+W_{\mathrm{SC}}\right)\right] -\sin\left[k_y\left(\frac{W+W_{\mathrm{SC}}}{2}\right) \right]\sinc\left(k_y\frac{W_{\mathrm{SC}}}{2}\right) \right\} ,\nonumber\\
f_2(k_y) &= \frac{1}{\wsc}\int^{W/2+W_{\mathrm{SC}}}_{W/2}dy \sin(k_y y)\nonumber\\
&=-\frac{1}{k_y\wsc}\left[\cos\left[k_y\left(\frac{W}{2}+W_{\mathrm{SC}}\right)\right]- \cos\left(k_y\frac{W}{2}\right)\right]\nonumber\\
&= \sin\left(k_y\frac{W_{\mathrm{SC}} +W}{2}\right)\sinc\left(k_y\frac{W_{\mathrm{SC}}}{2}\right).
\end{align}
For the case where $n\gg 1$, we have
\begin{align}
\Delta^{\pm}_{n} &=\Delta\frac{W_{\mathrm{SC}}}{2W_{\mathrm{SC}}+W}\left[(1+A_n)\cos\left(\frac{\phi}{2}\right)+\frac{E_Z}{\alpha k_{F_n}}\left(B_n\pm C_n\right) \sin\left(\frac{\phi}{2}\right)\right],
\end{align}
where
\begin{align}\label{eq:anbncn}
A_n &=   - \cos[\kfy(2W_{\mathrm{SC}}+W)]\cos\left[\kfy\left(W_{\mathrm{SC}}+W\right)\right]\sinc\left(\kfy W_{\mathrm{SC}}\right), \nonumber\\
B_{n} &= 2\cos^2\theta_n \sin\left[\kfy\left(2W_{\mathrm{SC}}+W\right)\right]  \sin\left[\kfy\left(\frac{W_{\mathrm{SC}}+W}{2}\right)\right] \sinc\left(\kfy\frac{W_{\mathrm{SC}}}{2}\right) \nonumber\\
&\hspace{0.2 cm} - 4 \frac{\alpha \kfy}{\hbar v_{F_n}} (2\wsc+W)\sin\theta_n  \sin\left[\kfy\left(2W_{\mathrm{SC}}+W\right) \right]\sin\left[\kfy\left(W_{\mathrm{SC}}+W\right)\right] \sinc(\kfy W_{\mathrm{SC}})\nonumber\\
&\hspace{0.2 cm} + 4\frac{\alpha}{\hbar v_{F_n}}\sin\theta_n\sin\left[\kfy(2W_{\mathrm{SC}}+W)\right]\sinc\left[\kfy\left(\frac{W_{\mathrm{SC}}}{2}\right)\right]\nonumber\\
&\hspace{0.2 cm}\times\left\{ \frac{\kfy W}{2}\cos\left[\kfy\left(\frac{W_{\mathrm{SC}}+W}{2}\right)\right] - \sin\left[\kfy\left(\frac{W+W_{\mathrm{SC}}}{2}\right) \right] + \sin\left[\kfy\left(W_{\mathrm{SC}} +\frac{W}{2}\right)\right]\right\},\nonumber\\
C_{n} &=  2  \sin(2\theta_n) \sin\left[\kfy\left(2W_{\mathrm{SC}}+W\right) \right]\sin\left[\kfy\left(W_{\mathrm{SC}}+W\right)\right] \sinc(\kfy W_{\mathrm{SC}})\nonumber\\
&\hspace{0.2cm} - \sin(2\theta_n)(1 + \cos[\kfy(2W_{\mathrm{SC}}+W)])\sin\left[k_{F_{n,y}}\left(\frac{W_{\mathrm{SC}}+W}{2}\right)\right] \sinc\left(k_{F_{n,y}}\frac{W_{\mathrm{SC}}}{2}\right)\nonumber\\
&\hspace{0.2 cm} -4 \frac{\alpha}{\hbar v_{F_n}} \left(1-\cos\theta_n\cos\left[\kfy\left(2W_{\mathrm{SC}}+W\right)\right]\right)\sinc\left[\kfy\left(\frac{W_{\mathrm{SC}}}{2}\right)\right]\nonumber\\
&\hspace{0.6 cm}\times\left\{  \frac{\kfy W}{2}\cos\left[\kfy\left(\frac{W_{\mathrm{SC}}+W}{2}\right)\right] - \sin\left[\kfy\left(\frac{W+W_{\mathrm{SC}}}{2}\right) \right] + \sin\left[\kfy\left(W_{\mathrm{SC}} +\frac{W}{2}\right)\right]\right\}.
\end{align}
In the limit where $\alpha k_{F_n}\ll \varepsilon_{F_n}, \hbar v_{F_n}/(2W_{\mathrm{SC}}+W)$, we can further simplify the above equations to
\begin{align}\label{eq:anbncnsimp}
A_n &=   - \cos[\kfy(2W_{\mathrm{SC}}+W)]\cos\left[\kfy\left(W_{\mathrm{SC}}+W\right)\right]\sinc\left(\kfy W_{\mathrm{SC}}\right), \nonumber\\
B_{n} &= 2\cos^2\theta_n \sin\left[\kfy\left(2W_{\mathrm{SC}}+W\right)\right]  \sin\left[\kfy\left(\frac{W_{\mathrm{SC}}+W}{2}\right)\right] \sinc\left(\kfy\frac{W_{\mathrm{SC}}}{2}\right),\nonumber\\
C_{n} &=  2  \sin(2\theta_n) \sin\left[\kfy\left(2W_{\mathrm{SC}}+W\right) \right]\sin\left[\kfy\left(W_{\mathrm{SC}}+W\right)\right] \sinc(\kfy W_{\mathrm{SC}})\nonumber\\
&\hspace{0.2cm} - \sin(2\theta_n)(1 + \cos[\kfy(2W_{\mathrm{SC}}+W)])\sin\left[\kfy\left(\frac{W_{\mathrm{SC}}+W}{2}\right)\right] \sinc\left(\kfy\frac{W_{\mathrm{SC}}}{2}\right).
\end{align}
So, the gap is given by
\begin{align}\label{eq:gapsuppl}
\Delta^{\pm}_{n} &= \widetilde{\Delta}_{n}^{\pm} + \delta \Delta_{n}^{\pm} \nonumber\\
&=\Delta\frac{W_{\mathrm{SC}}}{2W_{\mathrm{SC}}+W}\left[\left(1+A_n\right)\cos\left(\frac{\phi}{2}\right) + \frac{E_{Z,||}}{\alpha k_{F_n}}(B_{n} \pm C_{n}) \sin\left(\frac{\phi}{2}\right)\right].
\end{align}

Expanding the above equation around $\phi =\pi$, we have
\begin{align}
\Delta_{n}^{\pm}(\pi +\delta \phi_{n}^{\pm}) &\approx \Delta\frac{W_{\mathrm{SC}}}{2W_{\mathrm{SC}}+W}\left[-\frac{1}{2}\left(1+A_n\right)\delta\phi^{\pm}_n + \frac{E_{Z,||}}{\alpha k_{F_n}}(B_{n} \pm C_{n})\right].
\end{align}
To the first order in Zeeman energy, the gap-closing point moves away from $\phi = \pi$ by
\begin{align}
\delta \phi^{\pm}_{n} &= \frac{2}{(1+A_n)}\frac{E_{Z,||}}{\alpha k_{F_n}} (B_n \pm C_n).
\end{align}
So, the gap closing point for each band moves away from $\phi = \pi$, forming a line that is linear in the Zeeman energy $E_{Z,||}$ and inversely proportional to the Fermi momentum $k_{F_n}$.

The vanishing of gap at $\phi = \pi$ for $E_{Z,||} = 0$ follows from a more general argument that at this special point ($\phi = \pi$, $E_{Z,||} = 0$), the Hamiltonian respects the mirror and time-reversal symmetries. These two symmetries imply that
\begin{align}
\widetilde{F}^{\pm}_{n}(y) &= \widetilde{\varphi}^{\pm*}_{n,k_{F_{n,x}},\uparrow}(y) \widetilde{\varphi}^{\mp*}_{n,-k_{F_{n,x}},\downarrow}(y) - \widetilde{\varphi}^{\pm*}_{n,k_{F_{n,x}},\downarrow}(y) \widetilde{\varphi}^{\mp*}_{n,-k_{F_{n,x}},\uparrow}(y)\nonumber\\
&=|\widetilde{\varphi}^{\pm}_{n,k_{F_{n,x}},\uparrow}(y)|^2 + |\widetilde{\varphi}^{\pm}_{n,k_{F_{n,x}},\downarrow}(y)|^2\nonumber\\
&=|\widetilde{\varphi}^{\pm}_{n,k_{F_{n,x}},\downarrow}(-y)|^2 + |\widetilde{\varphi}^{\pm}_{n,k_{F_{n,x}},\uparrow}(-y)|^2\nonumber\\
&=\widetilde{F}^{\pm}_{n}(-y),
\end{align}
where in the second and third lines above, we use the relation between the spinors with their mirror and time-reversal partners [Eqs.~\eqref{eq:my} and~\eqref{eq:wavefuncT}], respectively. Since $\widetilde{F}^{\pm}_{n}(y)$ and $G_{nn}(y)$ are even functions of $y$ while at $\phi = \pi$, $\Delta(y) = i\Delta\sgn(y)\Theta(W_{\mathrm{SC}}+W/2-|y|)\Theta(|y|-W/2)$ is an odd function of $y$, the superconducting gap $\widetilde{\Delta}_{n}^{\pm} = \frac{2}{L_y}\int_{-L_y/2}^{L_y/2} \Delta(y) G_{nn}(y) \widetilde{F}^{\pm}_{n}(y)$ then vanishes at $E_{Z,||} = 0$ and $\phi = \pi$. This gap closing manifests as the point where the BDI-phase transition lines emerge in the ``fan"-shaped region of the BDI phase diagram [see Fig.~\ref{fig:phasediagram} of the main text].

\section{Three and Four Phase intersection points in the BDI phase diagram}\label{sec:phasebound}
\begin{figure}[h!]
\centering
{\includegraphics[width = \linewidth]{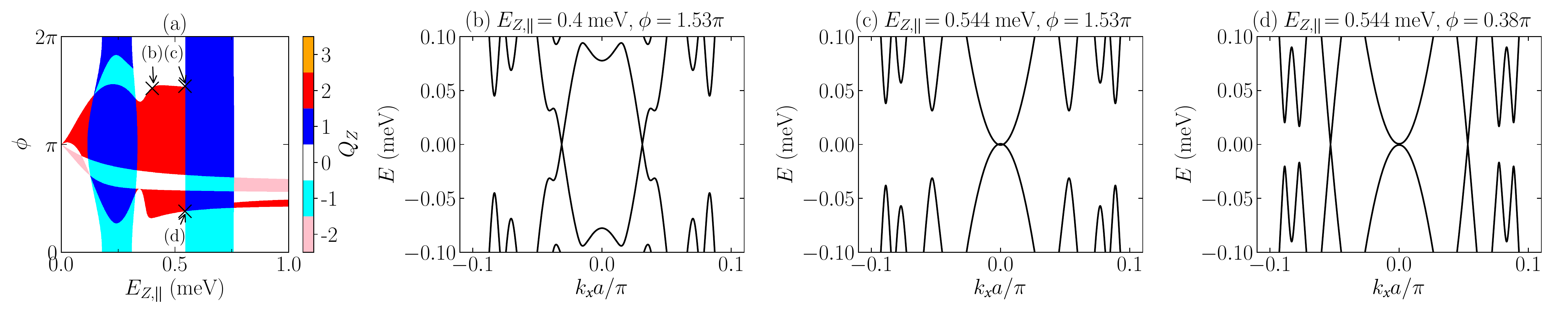}}
\caption{(a) Class BDI phase diagram and (b)-(d): Energy spectra for (b) a BDI phase transition point, (c) a three-phase intersection point, and (d) a four-phase intersection point. The parameters used are the same as those used in Fig.~\ref{fig:SNS_schematic} of the main text.}
\label{fig:spectrum_intersect}
\end{figure}

The BDI phase diagram [see Fig.~\ref{fig:spectrum_intersect}(a)] consists of phase boundaries points where three or four different phases meet. In this section, we elucidate the origin of these two situations. 
At points where four phases meet, two phase boundaries intersect: One separates phases whose $Q_{\mathbb{Z}}$ differ by $\pm 2$, and the other separates phases with different $Q_{\mathbb{Z}_2}$ indices. The former corresponds to a gap closing away from $k_x=0$, i.e., at $\pm k_{F_n}$ [see Fig.~\ref{fig:spectrum_intersect}(b)], whereas the latter corresponds to a gap closing at $k_x=0$ [see Fig.~\ref{fig:spectrum_intersect}(c)]. The intersection of these two phase boundaries give rise to either a three-phase intersection point with a gap closing at $k_x = 0$ [see Fig.~\ref{fig:spectrum_intersect}(c)] or a four-phase intersection point with a simultaneous gap closing at $k_x=0$ and $k_x=\pm k_{F_n}$ [see Fig.~\ref{fig:spectrum_intersect}(d)]. Note that the gap closings at $k_x=0$ and $k_x = \pm k_{F_n}$ do not necessarily occur in the same subband. In the two-parameter plane of $\phi$ and $E_{Z,\parallel}$ there are generically isolated points where the two transitions occur simultaneously giving rise to a four-phase intersection point in the phase diagram. 

To understand the origin of the three-phase intersection points, we examine the Hamiltonian projected onto a single subband $n$ near $k_x=0$. From symmetry consideration, the Hamiltonian has the form
\begin{equation}\label{eq:Hn}
\mathcal{H}_n = \left(\frac{\hbar^2k_x^2}{2 m^*_n} - \varepsilon_{F_n} (E_{Z,\parallel},\phi)\right) \rho_z + \delta_n (E_{Z,\parallel},\phi) k_x \rho_x,
\end{equation}
where $\varepsilon_{F_n}$ and $\delta_n$ are some functions of $\phi$ and $E_{Z,\parallel}$, and we assume that $m^*_n>0$ with $m^*_n$ being the effective electron mass for the $n$th subband. Here, $\rho_{x,y,z}$ are the Pauli matrices acting in the particle-hole basis $(\varphi_{n,k_x},[\varphi_{n,-k_x}]^\dagger)^T$ of the $n$th subband. Phase transitions in class D correspond to a sign change in $\varepsilon_{F_n}$. If $\varepsilon_{F_n}>0$, then sign changes of $\delta_n$ correspond to phase transitions where the class BDI index changes by $\pm 2$; in contrast, if $\varepsilon_{F_n}<0$, the subband is above the chemical potential, and a sign change in $\delta_n$ does not involve a gap closing. Consequently, at points where both $\varepsilon_{F_n}=0$ and $\delta_n=0$, a three-phase intersection point occurs. 

\section{Effects of breaking the BDI symmetry}\label{sec:breakBDI}
\subsection{Applying a transverse Zeeman field}\label{sec:transB}
\begin{figure}
\centering
{\includegraphics[width = \linewidth]{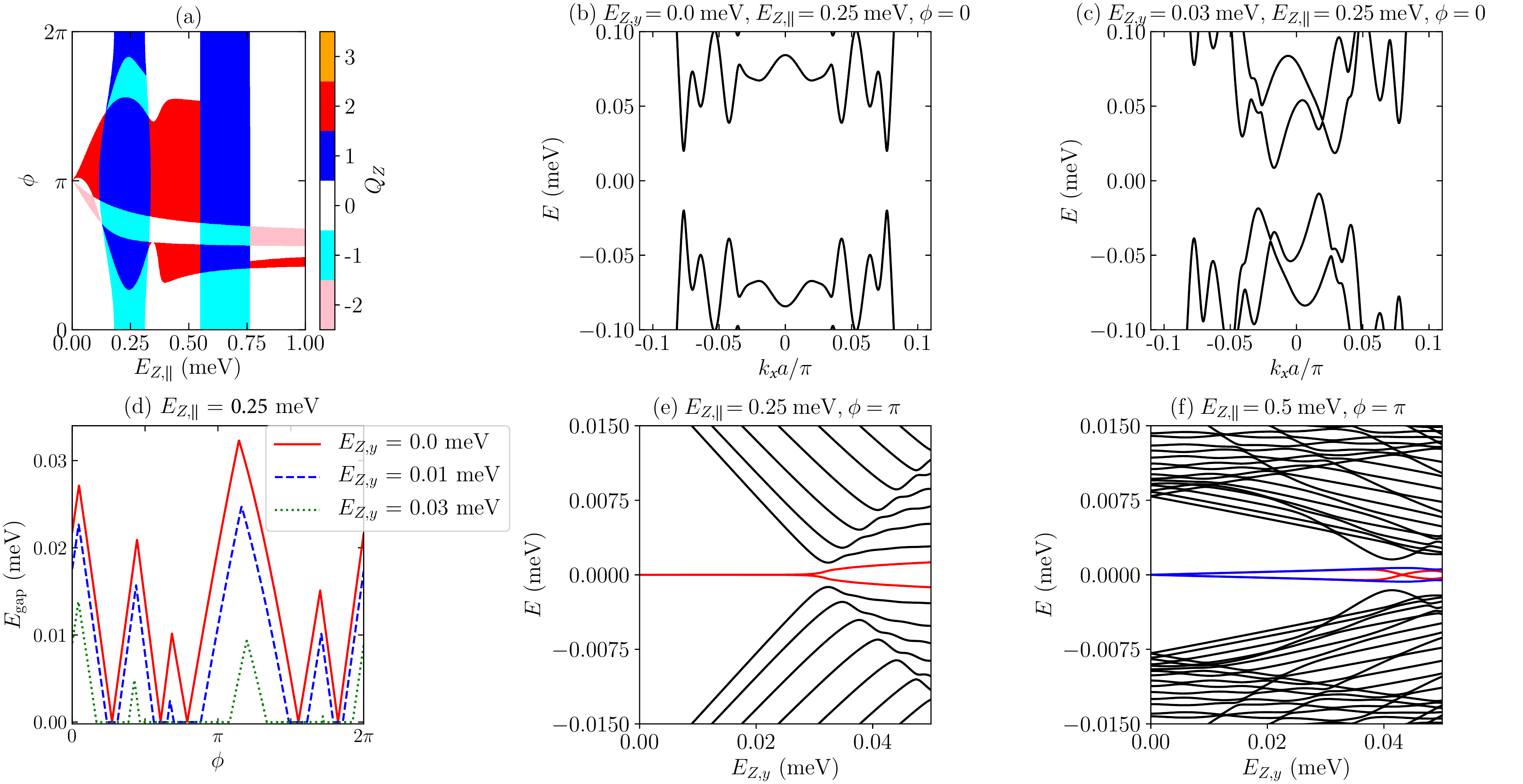}}
\caption{(a) A BDI phase diagram for a JJ with narrow leads in the absence of a transverse Zeeman field ($E_{Z,y} = 0$). (b),(c) Effect of a transverse Zeeman field on the energy spectrum. (b) Energy spectrum of the JJ for zero transverse field ($E_{Z,y} = 0$ meV), (c) Energy spectrum of the JJ for finite transverse field ($E_{Z,y} = 0.03$ meV). The transverse magnetic field tilts the energy spectrum, which reduces the spectral gap. (d) Effect of applying a transverse field on the bulk gap. The bulk gap [$E_{\mathrm{gap}} = \mathrm{min}_{k_x} E(k_x)$] along the $E_{Z,||} = 0.25$ meV linecut of the phase diagram in panel (a) calculated for increasing values of transverse magnetic field $E_{Z,y}$. When $E_{Z,y}$ = 0, the system is in the symmetry class BDI where each of the gap closings corresponds to a topological phase transition between phases whose BDI indices ($Q_{\mathbb{Z}}$) differ by $\pm 2$. When $E_{Z,y} \neq 0$, the $\widetilde{T}$ symmetry is broken and the spectral gap decreases, and gapless regions appear. (e)-(f) Effect of a transverse field on the Majorana zero modes at the end of a finite-length junction. (e) A single Majorana zero mode (red lines) at the each end of the junction stays at zero energy until the transverse Zeeman field is large enough to sufficiently reduce the gap such that the two MZMs from both ends overlap with each other. (f) In a region with $Q_{\mathbb{Z}}=2$, two Majorana zero modes (red and blue lines) at the same end of the junction hybridize and split in energy when a transverse Zeeman field is applied. The parameters used are the same as those used in Fig.~\ref{fig:SNS_schematic}(c) of the main text, i.e., $m^* = 0.026m_e$, $\mu = 0.6$ meV, $\alpha = 0.1$ eV{\AA}, $\Delta = 0.15$ meV [$\xi = \hbar v_F/(\pi \Delta) = 126$ nm], $W = 80$ nm, and $W_{\mathrm{SC}} = 160$ nm. Panels (a)-(d) are calculated for an infinitely long junction while panels (e)-(f) are calculated for $L_x = 30$ $\mu$m.}
\label{fig:Spectrum_Bny}
\end{figure}
One way to break the BDI symmetry is by applying a transverse in-plane Zeeman field $B_y$ perpendicular to the junction. The transverse Zeeman field tilts the band diagram [as shown in Fig.~\ref{fig:Spectrum_Bny}(c)] which decreases the spectral gap [Fig.~\ref{fig:Spectrum_Bny}(d)]. Applying the transverse field also breaks the effective ``time-reversal" symmetry of the system which changes the symmetry class from BDI into class D. As a result, Majorana end modes couple to one another and turn into finite energy modes, generically leaving either no or one zero-energy mode [see Fig.~\ref{fig:Spectrum_Bny}(f)], depending on the class D topological invariant being trivial or topological. Increasing the transverse Zeeman field strength further decreases the gap where beyond a certain value of Zeeman field, the two Majorana zero-modes from each end of the junction overlap with each other and split in energy [see Figs.~\ref{fig:Spectrum_Bny}(e) and~\ref{fig:Spectrum_Bny}(f)].

\subsection{Asymmetry in the left and right superconducting pairing potential}
\begin{figure}
\centering
{\includegraphics[width = 0.5\linewidth]{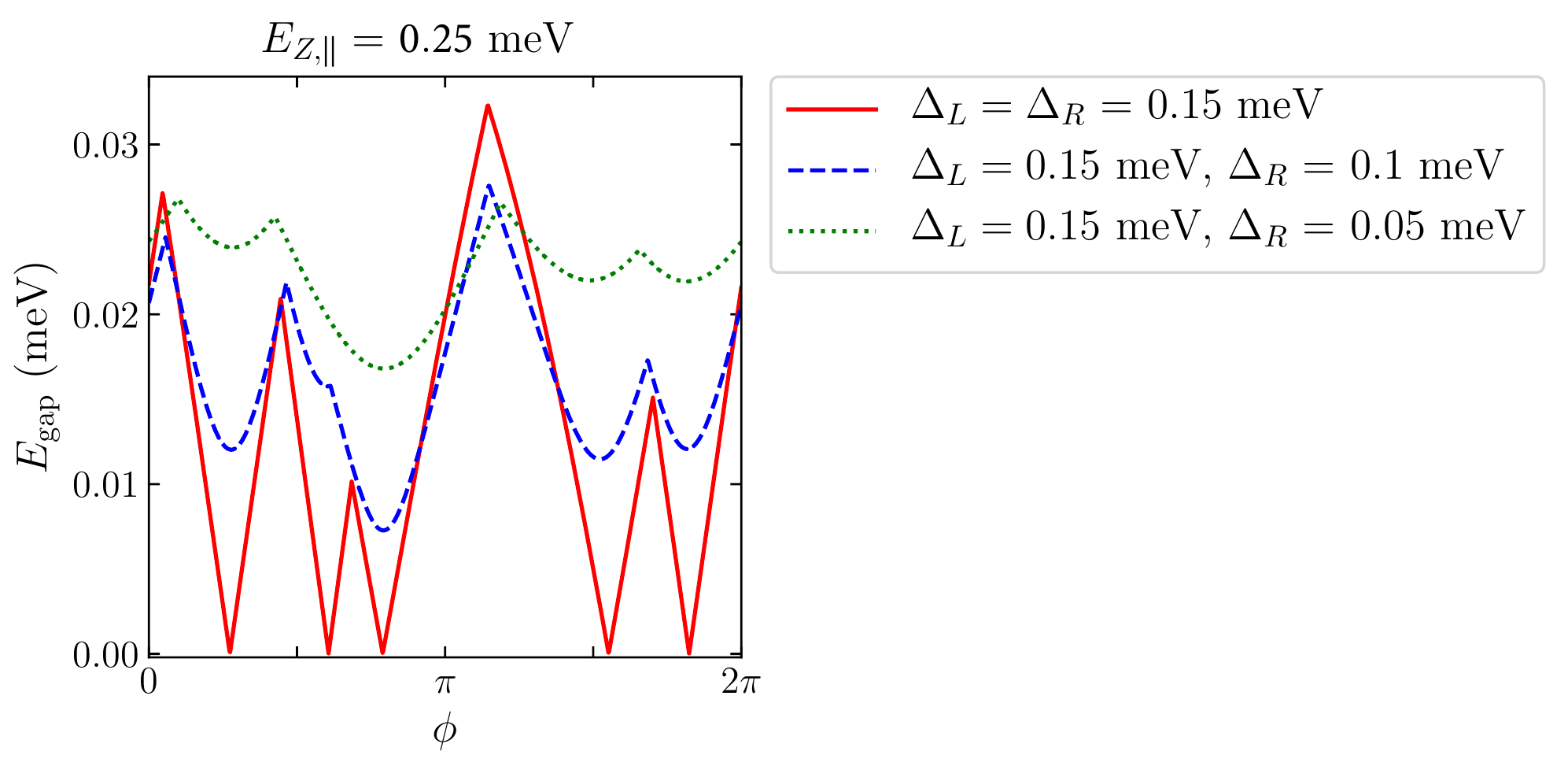}}
\caption{Effect of an asymmetry in the left and right superconducting pairing potentials on the spectral gap. The bulk gap [$E_{\mathrm{gap}} = \mathrm{min}_{k_x} E(k_x)$] along the $E_{Z,||} = 0.25$ meV linecut of the phase diagram in Fig.~\ref{fig:Spectrum_Bny}(a) calculated for increasing values of asymmetry between the left ($\Delta_L$) and right  ($\Delta_R$) pairing potential. When $\Delta_L = \Delta_R$, the system is in the symmetry class BDI where each of the gap closings corresponds to a BDI topological phase transition between phases whose BDI indices ($Q_{\mathbb{Z}}$) differ by $\pm 2$. Each of these gap closings happens at $k_x = k_{F_n}$ for different band. When $\Delta_{L} \neq \Delta_R$, the $\widetilde{T}$ symmetry is broken and the gap closing points are removed. The parameters used are the same as those used in Fig.~\ref{fig:SNS_schematic}(c) of the main text.}
\label{fig:gap_diffdelta}
\end{figure}

Similar to applying a transverse in-plane Zeeman field, an asymmetry in the left and right superconductor, i.e, different superconducting pairing potential ($\Delta_{L} \neq \Delta_R$) or different superconductor width also changes the symmetry class of the system from BDI to D. As a result, Majorana end modes at the same end of the junction couple to each other and splits into finite-energy modes. This leaves either zero or one Majorana mode at each end of the junction. However, contrary to the transverse in-plane Zeeman field, an asymmetry in the left and right superconductor lifts the gap closing points (which signify the BDI phase transitions) and generally increases the spectral gap as shown in Fig.~\ref{fig:gap_diffdelta}.

\section{Josephson Current}\label{sec:Josephson}

\subsection{Numerical calculation at finite temperature}
\begin{figure}
\centering
{\includegraphics[width = \linewidth]{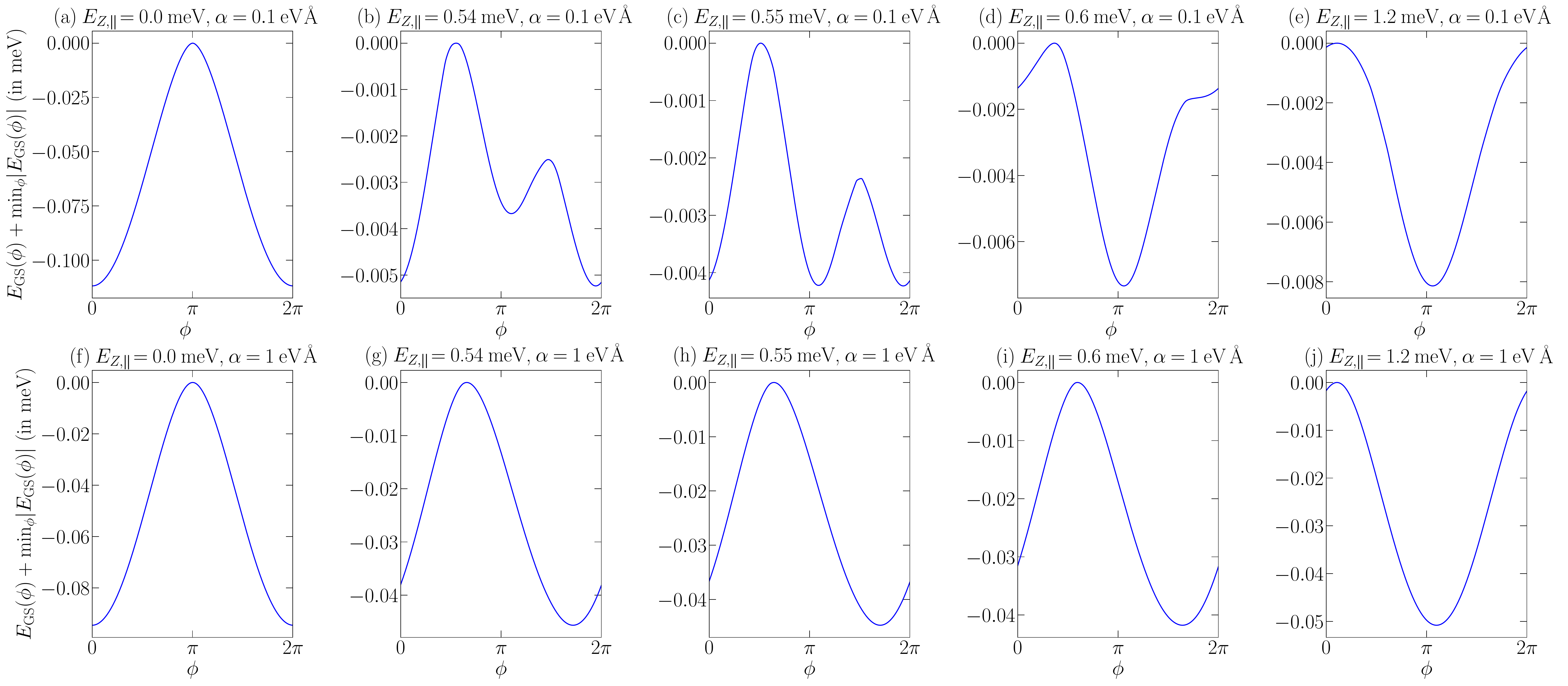}}
\caption{Ground-state energy $E_{\mathrm{GS}}$ versus superconducting phase difference $\phi$ of the two JJs whose Josephson currents are shown in Fig.~\ref{fig:Josephson}. The plots are given for increasing in-plane Zeeman fields $E_{Z,||}$ (from left to right) and different SOC strength. [(a)-(e)]: $\alpha = 0.1$ eV~\AA, [(f)-(j)]: $\alpha = 1$ eV~\AA. For small SOC strength [(a)-(e)], e.g., $\alpha = 0.1$ eV~\AA, the value of $\phi$ for which $E_{\mathrm{GS}}$ is minimized has an abrupt change from $\phi \approx 0$ to $\phi \approx \pi$ at the Zeeman field strength for which the minimum of the critical current occurs ($E_{Z,||} \approx 0.55$ meV) [see panel (c)]. This phase jump is a first-order phase transition. However, it does not correspond to a topological phase transition. The latter occurs at $E_{Z,\parallel} \approx 0.27$ meV [see Fig.~\ref{fig:phasediagram}(c)].  [(f)-(j)]: For large SOC strength, e.g., $\alpha = 1$ eV~\AA, the value of $\phi$ for which $E_{\mathrm{GS}}$ is minimized changes continuously [see Fig.~\ref{fig:Josephson}(b)]. The parameters used are the same as those used for Fig.~\ref{fig:Josephson}.}
\label{fig:groundstate}
\end{figure}

The Josephson current at temperature $T$ is computed by first calculating the many-body partition function which is given by
\begin{equation}
Z = \prod_j(1+e^{-\beta E_j}) = 2 \prod_j \cosh^2\frac{\beta E_j}{2},
\end{equation}
where $\beta = (k_BT)^{-1}$ and the product is evaluated over all energy states labeled by $E_j$. The free energy is
\begin{equation}\label{eq:freeenergy}
\mathcal{F} = - \frac{\mathrm{ln} Z}{\beta} = -\frac{4}{\beta}\sum_j \mathrm{ln} \left(\cosh\left( \frac{\beta E_j}{2}\right) \right).
\end{equation}
Figure~\ref{fig:groundstate} shows the numerically calculated ground-state energy $E_{\mathrm{GS}} = \mathcal{F} (T = 0)$ versus superconducting phase difference for a JJ with different Zeeman field strength. As shown in the upper panel of Fig.~\ref{fig:groundstate}, for small SOC strength $\alpha$ (e.g., $\alpha$ = 0.1 eV\AA), the ground state exhibits a $\pi$-phase jump at the Zeeman field strength ($E_{Z,||} = 0.55$ meV) where the minimum in the critical current happens (see Fig.~\ref{fig:Josephson} of the main text). This $\pi$-phase jump signifies a first-order phase transition and is associated with the minimum of the critical current in Fig.~\ref{fig:Josephson}(a). Since the critical field at which the minimum of the critical current happens ($E_{Z,||} = 0.55$ meV) is not the same as the critical field at which the topological phase transition happens  [$E_{Z,||} = 0.27$ meV, as shown in Fig.~\ref{fig:phasediagram}(c)], the minimum of the critical current does not indicate a topological phase transition. For large $\alpha$, the phase gradually shifts with the Zeeman energy $E_{Z,||}$ (see lower panel of Fig.~\ref{fig:groundstate}) instead of exhibiting an abrupt $\pi$-phase shift. Thus, the minima in the critical current for large $\alpha$ signify neither a first-order nor a topological phase transition.

The Josephson current can be calculated from the free energy [Eq.~\eqref{eq:freeenergy}] as
\begin{equation}\label{eq:Josephcurrent}
I(\phi) = \frac{2e}{\hbar}\frac{d\mathcal{F}}{d\phi} = -\frac{4e}{\hbar} \sum_{j} \mathrm{tanh}\left(\frac{\beta E_j}{2} \right) \frac{dE_j}{d\phi}.
\end{equation}
We use Eq.~\eqref{eq:Josephcurrent} to numerically calculate the Josephson current plotted in Fig.~\ref{fig:Josephson} of the main text.

\subsection{Analytical calculation at zero temperature}
To understand the Josephson current obtained from the full numerical calculation, in the following we are going to derive the analytical formula for the Josephson current. We focus only on the zero-temperature results. However, our results hold qualitatively also for the finite-temperature case. In the following we will work in two different limits: small SOC strength ($\alpha k_{F_n} \ll E_{Z,||}$) and  large SOC strength ($\alpha k_{F_n} \gg E_{Z,||}$). 

\begin{figure}[h!]
\centering
{\includegraphics[width = 0.6\linewidth]{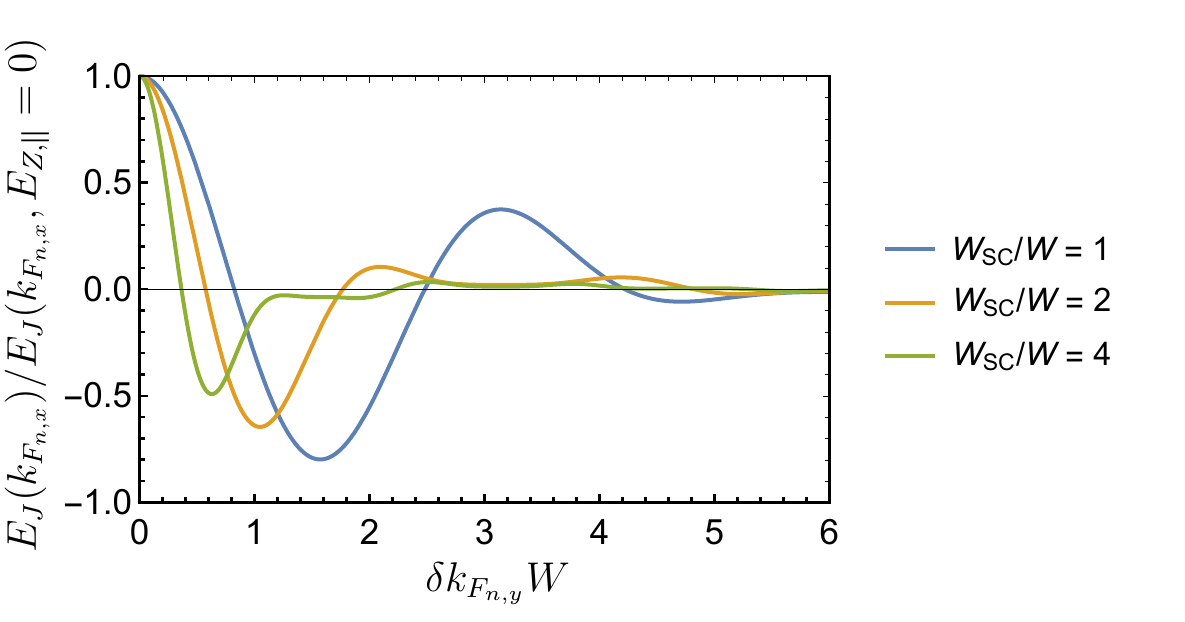}}
\caption{Plot of the Josephson coupling energy for a single-subband $E_{J}(k_{F_{n,x}})$ [Eq.~\eqref{eq:Ej}] as a function of $\delta k_{F_{n,y}}W$ for different values of superconductor width $W_{\mathrm{SC}}$.  
}
\label{fig:Integral}
\end{figure}

To explain the minimum in the critical current for small SOC strength ($\alpha k_{F_n} \ll E_{Z,||}$), for simplicity we take the limit where $\alpha = 0$ and $\Delta \ll E_{Z,||} \ll \varepsilon_{F_n}$. To this end, we first calculate the zero-temperature free energy. Since the
dominant contribution to the $\phi$-dependent part of the ground-state energy comes from Andreev bound states near the Fermi energy, we can write down the free energy as
\begin{equation}
\mathcal{F} = \text{const} - \sum_{n}^N E_J(k_{F_{n,x}}) \cos(\phi)
\end{equation}
where the Josephson energy is given by
\begin{align}\label{eq:Ejkx}
E_J(k_{F_{n,x}}) &= \Delta^2 \int_{-W/2-W_{\mathrm{SC}}}^{-W/2} dy \int^{W/2+W_{\mathrm{SC}}}_{W/2} dy' \int_0^\infty d\tau \left[ \langle \psi_{n,k_{F_{n,x}},\uparrow}^\dagger(y,\tau) \psi_{n,k_{F_{n,x}},\downarrow}^\dagger(y,\tau) \psi_{n,k_{F_{n,x}},\downarrow} (y',0)\psi_{n,k_{F_{n,x}},\uparrow}(y',0) \rangle + \mathrm{c.c.}\right] \nonumber\\
&= \Delta^2 \int_{-W/2-W_{\mathrm{SC}}}^{-W/2} dy \int^{W/2+W_{\mathrm{SC}}}_{W/2} dy' \int d\tau \left[ G_{n,k_{F_{n,x}},\uparrow}(y-y',\tau)G_{n,k_{F_{n,x}},\downarrow}(y-y',\tau)+ \mathrm{c.c.} \right].
\end{align}
Note that in the second line of Eq.~\eqref{eq:Ejkx}, we have used Wick's Theorem. To evaluate the above equation, we first linearize the field operator for spin $s = \uparrow/\downarrow$ near the Fermi momentum:
\begin{equation}
\psi_{n,k_{F_{n,x}},s}(y) \simeq \psi_{n,k_{F_{n,x}},s,+}(y) e^{ik_{F_{n,y},s}y} + \psi_{k_{F_{n,x}},s,-}(y) e^{-ik_{n,F_{n,y},s}y}.
\end{equation}
The Green's function is then given by
\begin{align}
G_{n,k_{F_{n,x}},s}(y-y',\tau) &= \langle \psi_{n,k_{F_{n,x}},s}(y,\tau) \psi^\dagger_{n,k_{F_{n,x}},s}(y',0) \rangle \nonumber\\
&= \langle \psi_{n,k_{F_{n,x}},s,+}(y,\tau) \psi^\dagger_{n,k_{F_{n,x}},s,+}(y',0) \rangle e^{ik_{F_{n,y,s}}(y-y')} +\langle \psi_{n,k_{F_{n,x}},s,-}(y,\tau) \psi^\dagger_{n,k_{F_{n,x}},s,-}(y',0) \rangle e^{-ik_{F_{n,y,s}}(y-y')}\nonumber\\
&= \frac{e^{ik_{F_{n,y,s}}(y-y')}}{y-y'-iv_{F_n}\tau} + \frac{e^{-ik_{F_{n,y,s}}(y-y')}}{y-y'+iv_{F_n}\tau} .
\end{align}
The time integral in Eq.~\eqref{eq:Ejkx} can be evaluated to be
\begin{align}\label{eq:intG}
\int d\tau &G_{n,k_{F_{n,x}},\uparrow}(y-y',\tau) G_{k_{F_{n,x}},\downarrow}(y-y',\tau) \nonumber\\
&= \int_0^\infty d \tau \frac{e^{i (k_{F_{n,y,\uparrow}} - k_{F_{n,y,\downarrow}}) (y-y')}}{(y-y')^2 + v_{F_n}^2 \tau^2} + \frac{e^{-i (k_{F_{n,y,\uparrow}} - k_{F_{n,y,\downarrow}}) (y-y')}}{(y-y')^2 + v_{F_n}^2 \tau^2}+ \frac{e^{i (k_{F_{n,y,\uparrow}} +k_{F_{n,y,\downarrow}}) (y-y')}}{[(y-y') -iv_{F_n} \tau]^2}  + \frac{e^{-i (k_{F_{n,y,\uparrow}} +k_{F_{n,y,\downarrow}}) (y-y')}}{[(y-y') +iv_{F_n} \tau]^2}  \nonumber\\
&= \frac{\pi}{v_{F_n}}  \left[\frac{e^{i\delta k_{F_{n,y}} (y-y')}}{y-y'} + \mathrm{c.c.} \right] - \frac{1}{i v_{F_n}} \left[ \frac{e^{2ik_{F_{n,y}}(y-y')} - e^{-2ik_{F_{n,y}}(y-y')}}{y-y'} \right],
\end{align}
where $\delta k_{F_{n,y}} =  k_{F_{n,y},\uparrow} - k_{F_{n,y},\downarrow} = 2E_{Z,||}/(\hbar v_{F_{n,y}})$ and $2k_{F_{n,y}} \equiv k_{F_{n,y},\uparrow} + k_{F_{n,y},\downarrow}$.
Since the last term in Eq.~\eqref{eq:intG} is highly oscillatory in $(y-y')$,  it can be neglected and the Josephson energy is then given by
\begin{align}\label{eq:EJkF}
E_J (k_{F_{n,x}})&= \Delta^2 \int_{-W/2-W_{\mathrm{SC}}}^{-W/2} dy \int^{W/2+W_{\mathrm{SC}}}_{W/2} dy' \left[\frac{e^{i\delta k_{F_{n,y}} (y-y')}}{|y-y'|} +\mathrm{c.c.}\right].
\end{align}
To evaluate the integral in Eq.~\eqref{eq:EJkF}, we first differentiate it with respect to $\delta k_{F_{n,y}}$:
\begin{align}
\frac{\partial E_J}{\delta k_{F_{n,y}}} &\simeq 2\pi i   \frac{\Delta^2}{v_{F_n}}\int_{-W/2-W_{\mathrm{SC}}}^{-W/2} dy \int^{W/2+W_{\mathrm{SC}}}_{W/2} dy' \left[ e^{i\delta k_{F_{n,y}} y} e^{-i\delta k_{F_{n,y}} y'} - \mathrm{c.c.}\right]\nonumber\\
&=2 \pi i  \frac{\Delta^2}{v_{F_n}} \left[\left.\frac{e^{i\delta k_{F_{n,y}} y}}{i\delta k_{F_{n,y}}}\right|^{-W/2}_{-W/2-W_{\mathrm{SC}}}\left.\frac{e^{-i\delta k_{F_{n,y}} y'}}{-i\delta k_{F_{n,y}}}\right|^{W/2+W_{\mathrm{SC}}}_{W/2} - \mathrm{c.c.} \right] \nonumber\\
&=16 \pi  \frac{\Delta^2}{v_{F_n}} \sin[\delta k_{F_{n,y}}(W+W_{\mathrm{SC}})] \left[ \frac{\sin \left(\delta k_{F_{n,y}} W_{\mathrm{SC}}/2\right)}{\delta k_{F_{n,y}}} \right]^2.
\end{align}
So, we have the Josephson coupling energy as
\begin{align}\label{eq:Ej}
E_J(k_{F_{n,x}}) = 16 \pi\frac{\Delta^2}{v_{F_n}}\int d\delta k_{F_{n,y}} \sin[\delta k_{F_{n,y}} (W+W_{\mathrm{SC}})] \left[\frac{\sin \left( \delta k_{F_{n,y}} W_{\mathrm{SC}}/2\right)}{\delta k_{F_{n,y}}} \right]^2.
\end{align}
The Josephson current due to a single subband can then be calculated from
\begin{align}
I(k_{F_{n,x}},\phi) = \frac{2e}{\hbar} E_J (k_{F_{n,x}}) \sin \phi.
\end{align}
The critical current is the sum of the contribution from all occupied subbands:
\begin{align}
I_c &= \max_{\phi} \sum_n I(k_{F_{n,x}},\phi)\nonumber\\
&= \frac{2e}{\hbar} \left|\sum_n E_J(k_{F_{n,x}})\right|.
\end{align}
Figure~\eqref{fig:Integral} illustrates the behavior of the contribution of a single subband to the Josephson energy as a function of $E_{Z,\parallel}$. The dependence on the band index originates from $\delta k_{F_{n,y}} = 2E_{Z,\parallel}/(\hbar v_{F_{n,y}})$. 

Having derived the Josephson current for small $\alpha$, in the following we are going to derive the Josephson current in the limit where $\alpha k_{F_n} \gg E_{Z,||}$. To obtain the Josephson current, we first derive the formula for the condensation energy,  which is the difference between the normal and superconducting ground-state energies. Linearizing the dispersion near the Fermi momentum, we have $E = \hbar v_{F_{n}}k_x$. The zero-temperature condensation energy is then given by

\begin{align}\label{eq:Econdsuppl}
E_{\mathrm{cond}} &= -\sum_n^N \sum_{m = \pm}\int_{-k_{F_{n}}}^{k_{F_{n}}} \frac{dk_x}{2k_{F_{n}}}  \left(\sqrt{(\Delta_{n}^m)^2 + (\hbar v_{F_{n}} k_x)^2} - \hbar v_{F_{n}} |k_x|\right) \nonumber\\
&= -\sum_n^N \sum_{m =\pm}\int_{-k_{F_{n}}}^{k_{F_{n}}} dk_x  \frac{\hbar v_{F_{n}}}{2k_{F_{n}}} |k_x| \left(\sqrt{1+\frac{(\Delta_{n}^m)^2}{(\hbar v_{F_n}k_x)^2}} - 1 \right) \nonumber\\
&\sim -\sum_n^N\sum_{m=\pm}\frac{(\Delta_{n}^m)^2}{\varepsilon_{F_{n}}} \int_{\Lambda}^{\frac{\hbar v_{F_{n}}k_{F_{n}}}{\Delta_{n}^m}} d \tilde{k}_x \tilde{k}_x\left(\sqrt{1+\frac{1}{\tilde{k}_x^2\color[rgb]{0,0,0}}} - 1 \right) \nonumber\\
&= -\sum_n^N\sum_{m =\pm}\frac{(\Delta^m_{n})^2}{\varepsilon_{F_{n}}} \log \left(\frac{1}{\Lambda} \frac{\varepsilon_{F_{n}}}{\Delta^m_{n}} \right),
\end{align}
where $\varepsilon_{F_n} = \hbar v_{F_n}k_{F_n}$ is the Fermi energy of $n$th band, $2N$ is the number of occupied subbands and $\Lambda$ is the low energy cutoff introduced in the integration. Substituting Eq.~\eqref{eq:gapsuppl} into Eq.~\eqref{eq:Econdsuppl} and ignoring the logarithmic dependence on the superconducting gap in Eq.~\eqref{eq:Econdsuppl}, we obtain the condensation energy to the leading order in $E_{Z,\parallel}$ as 
\begin{align}
E_{\mathrm{cond}} 
= -\left(\Delta\frac{W_{\mathrm{SC}}}{W+2W_{\mathrm{SC}}}\right)^2\sum_{n}^N\frac{1}{2\varepsilon_{F_{n}}} \left[(1+A_n)^2 [1+\cos \left(\phi\right)] - \sin(\phi)\zeta_n \right],
\end{align}
where
\begin{equation}
\zeta_{n} = \frac{E_{Z,||}}{\alpha k_{F_n}}\frac{B_n}{(1+A_n)},
\end{equation}
with $A_n$ and $B_n$ being given by Eq.~\eqref{eq:anbncn}. 

The Josephson current at zero temperature is given by $I(\phi) = \frac{2e}{\hbar}\frac{dE_{\mathrm{cond}}}{d\phi}$. 
So, in the limit where $E_{Z,||} \ll \alpha k_{F_n}$, the Josephson current changes gradually with the Zeeman field $E_{Z,||}$. In particular, the value of $\phi$ that minimizes the energy shifts linearly with $E_{Z,||}$.

\end{document}